\documentclass[modern, trackchanges]{aastex631}

\usepackage{xcolor}
\usepackage{soul}
\usepackage{ulem}

\newcommand{\sptran}{L7$-$T3}

\newcommand{\nsp}{325}
\newcommand{\nsamp}{270}
\newcommand{\nsbsp}{13}
\newcommand{\nsb}{8}
\newcommand{\nyoungsp}{51}
\newcommand{\nyoung}{49}
\newcommand{\nbenchsp}{21}
\newcommand{\nbench}{15}
\newcommand{\nyoungbench}{4}
\newcommand{\nstrong}{51}

\newcommand{\nvarcand}{62}
\newcommand{\nweak}{11} 
\newcommand{\sbcontamweak}{2.3}
\newcommand{\sbcontamstr}{2.3}

\newcommand{\nsynthbin}{10731}

\submitjournal{ApJ}

\shorttitle{Binaries or Variables?}
\shortauthors{Ashraf et al.}

\graphicspath{{./}{}}

\begin{document}

\title{Disentangling the Signatures of Blended-Light Atmospheres in L/T Transition Brown Dwarfs}

\correspondingauthor{Afra Ashraf}
\email{aa4085@barnard.edu}

\author[0000-0003-4912-5229]{Afra Ashraf}
\affiliation{Department of Physics \& Astronomy, Barnard College \\
3009 Broadway, New York, NY 10027, USA}
\affiliation{Department of Astrophysics, American Museum of Natural History \\
200 Central Park West, New York, NY 10024, USA}

\author[0000-0001-8170-7072]{Daniella C. Bardalez Gagliuffi}
\affiliation{Department of Astrophysics, American Museum of Natural History \\
200 Central Park West, New York, NY 10024, USA}

\author[0000-0003-0192-6887]{Elena Manjavacas}
\affiliation{AURA for the European Space Agency (ESA), ESA Office, Space Telescope Science Institute, 3700 San Martin Drive, Baltimore, MD, 21218 USA}

\author[0000-0003-0489-1528]{Johanna M. Vos}
\affiliation{Department of Astrophysics, American Museum of Natural History \\
200 Central Park West, New York, NY 10024, USA}

\author{Claire Mechmann}
\affiliation{Department of Physics \& Astronomy, Lehman College, City University of New York \\
250 Bedford Park Blvd W, The Bronx, NY 10468, USA}

\author[0000-0001-6251-0573]{Jacqueline K. Faherty}
\affiliation{Department of Astrophysics, American Museum of Natural History \\
200 Central Park West, New York, NY 10024, USA}



\begin{abstract}
We present a technique to identify spectrophotometrically variable \sptran~brown dwarfs with single-epoch, low-resolution, near-infrared SpeX spectra. We calculated spectral indices on known variable brown dwarfs and used them to select 11 index-index parameter spaces where known variables can be distinguished from the rest of the general population of brown dwarfs. We find \nvarcand~candidate variables, 12 of which show significant variability amplitude in independent photometric monitoring surveys. This technique constitutes the first formal method to identify a time-dependent effect such as variability from peculiarities in their integrated light spectra. This technique will be a useful tool to prioritize targets for future photometric and spectroscopic monitoring in the era of \textit{JWST} and 30\,m-class telescopes. 
\end{abstract}




\section{Introduction} 
\label{sec:intro}
Substellar variability is a time-dependent phenomenon caused by an inhomogeneous atmosphere over the course of an object’s rotation period, and it is characterized by a change in flux over time. Photometric variability is dominated by magnetic activity in M dwarfs and early-L dwarfs \citep{1993PASP..105..955H} and by inhomogenous cloud coverage in cooler brown dwarfs \citep{2002ApJ...577..433G}. The cool temperatures of L and T-type brown dwarf atmospheres permit the condensation of different species, resulting in patchy layers of large-scale cloud and haze structures \citep{2001ApJ...556..872A}.  Different elements and compounds reach their condensation equilibrium at varying temperatures and pressures, such that clouds can form at different altitudes in the atmosphere. Silicate and iron condensates begin to form in the atmospheres of L-dwarfs below 2400\,K \citep{2006asup.book....1L}. Towards the end of the L sequence, silicate and iron clouds sink beneath the photosphere in what is known as ``rain out'' \citep{2004AJ....127.3553K}. By the time an object has cooled down to T dwarf temperatures, their clouds have sunken below the photospheres. The sudden increase in brightness in the $J$-band for late-L/early-T brown dwarfs (called the ``J-band bump'') has been interpreted as evidence for this process~\citep{2002ApJ...571L.151B,2002AJ....124.1170D,2003AJ....126..975T,2004AJ....127.2948V}. An alternative explanation for photometric variability is a cloudless model, where variations arise due to a convective atmosphere dominated by CO/CH$_4$ radiative convection as shown in \citet{2015ApJ...804L..17T,2016ApJ...817L..19T,2017ApJ...850...46T,2019ApJ...876..144T,2020A&A...643A..23T}.


The L/T transition is a critical phase in the evolution of brown dwarfs which happens over a relatively short range of effective temperatures ($1150\, \text{K} \lesssim T_{eff} \lesssim 1350\, \text{K}$; \citealt{2019BAAS...51c.253V,2015ApJ...810..158F}) and is characterized by the appearance of methane in near-infrared (NIR) spectra and a shift towards blue in $J-H$ and $H-K$ colors. Strong NIR variability is most common within this spectral type range, with a 24\% variability occurrence (amplitudes $>2\%$) compared to only 3.2\% outside of the L/T transition \citep{2014ApJ...797..120R}. Since variability is thought to be more prevalent in this spectral type region, we have the best chance of identifying suitable targets amenable for further study.

However, variability studies are extremely resource intensive. Reliable light curves require precision photometric monitoring over the course of several hours for each target, covering a significant portion of an object's rotation period. For example, in their photometric variability survey, \citet{2014ApJ...793...75R} obtained 62 light curves of L4$-$T9 dwarfs with $2-3$\,m aperture ground-based telescopes spanning roughly 2-5 hours per target. Accounting for telescope slewing, target acquisition and data read out, these observations amounted to almost 250 hours of on-sky time, resulting in variability detection in 16\% of their targets. In their space-based variability survey, \citet{2015ApJ...799..154M} targeted 44 L3$-$T8 brown dwarfs and identified 19 objects, i.e. $61^{+17}_{-20}\%$ of L dwarfs and $31^{+35}_{-17}\%$ of T dwarfs, as having photometric variability with an amplitude of $0.2-4.6$\% using the \textit{Spitzer Space Telescope}. Likewise, \citet{2014ApJ...782...77B} found 6 variable brown dwarfs in a sample of 22 observed with the \textit{Hubble Space Telescope}. These surveys are crucial to understand the occurrence of variability and to study the time-dependent atmospheric dynamics and vertical structure of brown dwarf atmospheres at a population level, yet such surveys cannot pre-select candidates based on their likelihood to show variability because currently there is no systematic way to prioritize candidates for variability studies. 

Spectral binaries may hold a clue for such a technique. Spectral binaries are systems of two brown dwarfs of different spectral types with a combined peculiar unresolved, low-resolution, NIR spectrum \citep{2004ApJ...604L..61C,2007ApJ...659..655B, 2010ApJ...710.1142B, 2014ApJ...794..143B}. Spectral binary candidates are identified using spectral indices and binary template fitting, and later confirmed through high resolution imaging, radial velocity or astrometric variability~\citep[e.g.,][]{2015AJ....150..163B,2016ApJ...827...25B,2020MNRAS.495.1136S}. However, a handful of spectral binary candidates have not shown signs of binarity in high resolution follow-up and instead are known to show photometric variability. The young early-T dwarfs 2MASS J21392676+0220226~\citep[hereafter J2139+0220,][]{2006ApJ...637.1067B} and 2MASS J13243553+6358281~\citep[hereafter J1324+6358,][]{2007AJ....134.1162L}, were originally suggested to be spectral binary candidates~\citep{2010ApJ...710.1142B} of L4 and T5 components, but are actually single, photometrically variable brown dwarfs~\citep{2012ApJ...750..105R, 2015ApJ...799..154M}. This suggests that the spectral binary technique may also be inadvertently sensitive to the type of inhomogeneous atmospheres that drive variability. 

\citet{2013AJ....145...71K} searched for photometric variability in $J$ and $K$-bands using the Gemini infrared camera on the 3-meter Shane Telescope at Lick Observatory for 15 L and T-type brown dwarfs, including 7 spectral binary candidates. They found variability in 4 objects, two of which were spectral binary candidates at the time,  2MASS J2139+0220 and SDSS J141624.08+134826.7A~\citep[hereafter J1416+1348A,][]{2010AandA...510L...8S}. In their spectroscopic variability program with \textit{Hubble Space Telescope’s} Wide Field Camera 3, \citet{2019AJ....157..101M} used the spectral binary technique to identify peculiar sources and multiple systems. However, they discovered that five of the eight spectral binary candidates they found had been previously identified in the literature as photometrically variable. Sources like 2MASS~J2139+0220, 2MASS~J1324+6358, and SDSS J1416+1348A, while originally selected as spectral binary candidates, present no compelling signs of binarity and instead show significant variability amplitudes. These motivating examples suggest that the spectral signatures caused by two separate sources at different temperatures on an unresolved spectrum are similar to those of a single object with multiple cloudy, patchy layers at different temperatures. The peculiarities in the spectra selected by index-based spectral binary techniques may instead be occasionally caused by cloud-driven variability. 

Cloud-driven brown dwarf variability is also correlated with age. Young brown dwarfs have redder near-infrared and mid-infrared colors in comparison to field objects of the same spectral type. This is likely due to their low surface gravity and extended radii from their ongoing contraction, leading to slower sedimentation efficiency of scattering particles like dust grains in their atmospheres  \citep{2016ApJS..225...10F}. In the \textit{Weather on Other Worlds} survey by \citet{2015ApJ...799..154M}, they found an association with 92\% confidence between low surface gravity and high-amplitude variability among L3 - L5.5 dwarfs. \citet{2019MNRAS.483..480V} calculated a variability occurrence rate in a sample of L0-L8.5 young, low-gravity objects using the New Technology Telescope (NTT) and the United
Kingdom InfraRed Telescope (UKIRT) of $30_{-8}^{+16}\%$ which they compared to the results of \citet{2014ApJ...793...75R} where the field dwarfs have a variability occurrence rate of $11_{-4}^{+13}\%$. These results suggest that variability occurs more frequently and has a higher amplitude in young brown dwarfs. \citet{2012ApJ...754..135M} suggests that this correlation could be caused by high altitude clouds in low-gravity atmospheres, increasing the contrast between different cloud layers. In a recent variability survey, \citet{2022ApJ...924...68V} indeed finds that the maximum variability amplitudes observed for low-gravity L dwarfs are higher compared to field objects.

Variability studies in brown dwarfs are precursors to future studies of exoplanet climates. Brown dwarfs are characterized by physical properties that are very similar in nature to extrasolar planets, in particular directly-imaged exoplanets or young giant planets. L and T-type dwarfs overlap in temperature range ($500\, \text{K}$ $\lesssim T_{eff} \lesssim 2300\, \text{K}$;~\citealt{2015ApJ...810..158F}) with young giant exoplanets ($1300\, \text{K} \lesssim T_{eff} \lesssim 3000\, \text{K}$; \citealt{2014prpl.conf..739M}), a parallel reflected in their atmospheric spectra. However, brown dwarfs are much easier to image than exoplanets since they are typically unobstructed by a host star, making atmospheric studies more accessible. \citet{2016ApJ...820...40A} and \citet{2021MNRAS.503..743B} probed the giant exoplanets in the HR8799 system for variability using VLT/SPHERE, but neither could achieve high enough sensitivity to confidently detect variability. This highlights the importance of studying isolated planetary-mass objects as analogs to exoplanets in orbit around a bright host star.

The emergence of next-generation telescopes, such as the \textit{James Webb Space Telescope (JWST)}, will permit detailed characterization of exoplanets \citep{2021MNRAS.501.1999C}, and the techniques used to study brown dwarf atmospheres will be important predecessors to surface mapping and characterizing exoplanet atmospheres. We can analyze light curves to obtain variability amplitudes and rotation periods. By using stellar and substellar light curves, we can map the surface inhomogeneities of stars with topological light curve inversion and modeling \citep{2017Sci...357..683A,2019AJ....157...64L}, albeit with degeneracies. The same techniques will be applied to exoplanets with time-resolved high-contrast imaging. Our technique potentially offers an opportunity to probe the vertical structure of the atmosphere to complement the surface information provided by light curves and compare to 3D global circulation models~\citep{2019ApJ...883....4S,2021MNRAS.502.2198T}. 

In this paper we present a new technique to identify high likelihood variability candidates among \sptran~dwarfs based on spectral indices of single-epoch, low resolution, NIR spectra. In Section~\ref{sec:spex}, we discuss the characteristics of our sample. In Section~\ref{sec:candidates}, we use our technique to identify variability candidates for follow-up photometric monitoring. In Section~\ref{sec:discussion} we discuss the stability of SpeX as an instrument, calculate our variability fractions, and examine the possible surface features that could cause different variability amplitudes. Finally in Section~\ref{sec:conclusions} we present our conclusions.

\section{SpeX Spectral Sample}\label{sec:spex}

Our spectral sample was retrieved from the SpeX Prism Library (SPL;~\citealt{2014ASInC..11....7B}) and is composed of \nsp~spectra from \nsamp~unique objects within the \sptran~dwarf spectral type range ($1150\, \text{K} \lesssim T_{eff} \lesssim 1350\, \text{K}$). The SPL contains over 2000 NIR spectra of ultracool and brown dwarfs collaboratively collected since the year 2000 with SpeX, a spectrograph mounted on the 3-m NASA Infrared Telescope Facility in Mauna Kea, HI~\citep{2003PASP..115..389V, 2003PASP..115..362R}. We used spectra in the SPL acquired with the prism-dispersed mode of SpeX, which samples wavelengths between 0.75 and 2.5$\mu$m at a dispersion of 20–30\,\AA\,pixel$^{-1}$. The spectra contained in the SPL have been homogeneously obtained with the same observing strategy (see e.g., \citealt{2014ApJ...794..143B, 2019ApJ...883..205B, 2010ApJ...710.1142B} for technical details on acquisition, integration, calibration, and reduction of SpeX spectra).
 
Since we are interested in identifying differences between variable and non-variable objects across spectral features, such as a peculiar slope or depth of absorption lines and bands, we restrict our sample to spectra acquired in a single epoch. This means that we are not using spectra intended for spectroscopic variability, which involves a series of short-exposure spectra in a large field of view including reference stars to evaluate the change of spectral features over time.  Depending on how well the slit was aligned with the parallactic angle to avoid slit losses, and the duration of integration at the time of observations, the flux of an individual spectrum may fluctuate between 1-2\% of the total flux on separate epochs because of atmospheric turbulence (M. Connelley, priv. comm.). 

\begin{figure}
    \centering
    \includegraphics[width=\textwidth]{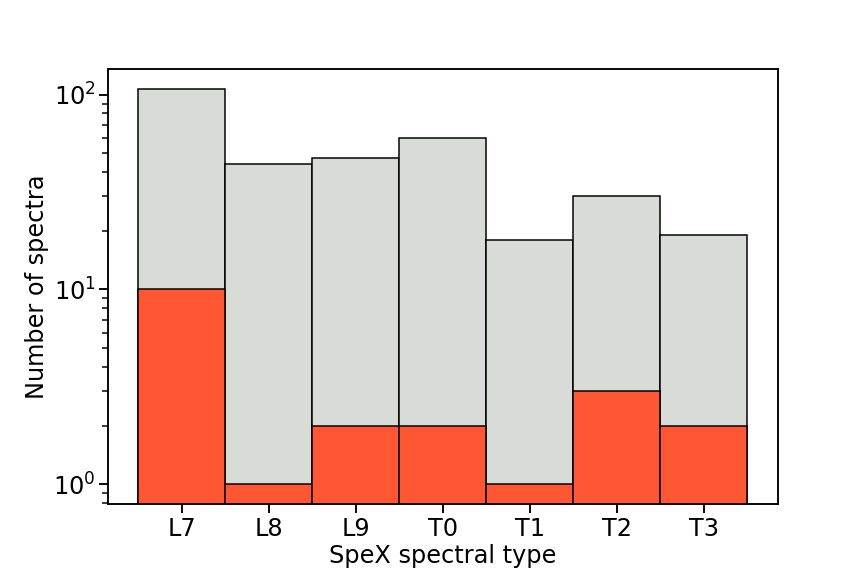}
    \caption{\textit{Spectral type distribution of our sample with benchmark variables in red.}}
    \label{fig:samplehist}
\end{figure}

We use the SpeX spectral classifications from the SPL, which follow the~\citet{2010ApJS..190..100K} standards and method that only use the $J$-band for spectral classification for L-dwarfs and the~\citet{2006ApJ...637.1067B} near-infrared spectral type standards for T-dwarfs. We excluded subdwarfs and sources in star-forming regions as we wanted to focus on differences in spectra caused by cloud-driven variability, not by other factors like reddening or low metallicity that would result in unusual spectra. The spectra were calibrated to absolute flux densities using their 2MASS $J$-band magnitudes and the empirical relations of~\citet{2015ApJ...810..158F}, with the flux calibration function within the \textit{\href{https://github.com/aburgasser/splat}{SPLAT}} package. 

Given the correlation between variability and age, we identified young objects within our sample following the literature. Of the \nsp~spectra examined, there are a total of \nyoungsp~young spectra from \nyoung~young objects, of which \nyoungbench~are confirmed variables from photometric monitoring. We consider objects as “young” according to the literature, given spectral features and kinematics significantly different from those in field objects, which are evident up to roughly 150\,Myr old (e.g., the age of the AB Doradus moving group, \citealt{2016ApJS..225...10F}). Objects with high space velocities relative to the Sun tend to have older ages, as they have had a long time to accumulate energy kicks from gravitational interactions with other stars on their orbit around the galactic center. Gravity estimates can be obtained from spectral features in low resolution spectra \citep{2013ApJ...772...79A} and age estimates by kinematic association to a young moving group, which like the Pleiades, AB Doradus, or the Hyades, are coeval, gravitationally-bound groups of stars. Additionally, we identify \nsbsp~spectra from \nsb~confirmed spectral binaries. The spectral type distribution of our final sample is shown in Figure~\ref{fig:samplehist}, highlighting variable sources. Since the SPL is a general repository of spectra observed for a variety of science cases, we do not claim our selected sample to be unbiased or complete.




\section{Identification of Candidate Variables}\label{sec:candidates}

\subsection{Benchmarks}\label{sec:bench}


From our sample of L/T transition objects, we select as ``benchmarks" \nbenchsp~spectra from \nbench~objects that have confirmed light curve amplitudes $>1\%$ in $J$, [3.6], or [4.5] bands from previous time-resolved, photometric monitoring across several variability studies varying in sensitivity and/or wavelength coverage (see Table~\ref{tab:benchmarks} and~\citealt{2020AJ....160...38V} for a summary of sources with measured variability). Ground-based near-infrared surveys \citep[e.g.][]{2014ApJ...793...75R,2019MNRAS.483..480V} are generally sensitive to variations greater than $1-2\%$, and can only observe each target as long as it is visible. Variability studies that use the WFC3 instrument on the Hubble Space Telescope \citep[e.g.][]{2013ApJ...768..121A,2014ApJ...782...77B} reach precisions of $0.01-0.02\%$, but have generally been limited in observing duration. Finally, variability studies using the Spitzer Space Telescope \citep[e.g.][]{2015ApJ...799..154M,2022ApJ...924...68V} achieve both sub-percent precision and relatively long monitoring duration ($\sim20~$hr). Variability monitoring with the Spitzer Space Telescope is likely the most sensitive to date for detecting variability in L and T brown dwarfs, yielding lower mid-IR amplitudes when compared to near-IR amplitudes. Our strict benchmark selection process discounted 8 objects from the literature with either upper limits in variability amplitudes (i.e., no variability or variability below the sensitivity limit) or previous identification as spectral binary candidates from the benchmark list that were later recovered as variability candidates: SDSS J0758+3247, SDSS J0858+3256, 2MASS J1119-1137, SDSS J1219+3128, SDSS J1254-0122, SDSS J1516+3053, 2MASS J1632+1904, 2MASS J2224-0158. This was also the case for Luhman 16A, which we decided to exclude from the benchmarks list due to contrasting reports of variability in the NIR and MIR.

The variability amplitudes in the NIR for our benchmarks range from 1.2\% to 26\% for 2MASS J21392676+0220226~\citep{2012ApJ...750..105R} in $J$-band, including Luhman 16B~\citep[11\% in WFC3-IR at 1.1-1.66 $\mu$m][]{2015ApJ...798..127B}. Three of our benchmarks only have variability measurements in~\textit{Spitzer} [3.6] or [4.5] bands with amplitudes between 1-3\%. While 2MASS J20025073-0521524 and  2MASSW J0310599+164816 do not have a measured rotation period, the rest of our benchmarks have periods between 2-19\,h. While the planetary-mass, L7 dwarf PSO 318.5-22~\citep{2013ApJ...777L..20L} qualifies as a benchmark given our criteria, we did not have a SpeX spectrum of this source in the SPL. 
Sources that had been previously selected as spectral binary candidates and for which no high-resolution imaging is available were excluded. This cut affected SDSS J075840.33+324723.4 and SDSS J151643.01+305344.4, both of which were identified as variability candidates with our technique. We excluded the blended light spectrum of Luhman 16AB from our benchmark list, but kept the individual spectrum of the spatially resolved B component. The B component is a T0 dwarf with a large variability amplitude of $\sim$ 11\%~\citep{2013AandA...555L...5G}, which gets diminished in the combined-light spectrum. We also excluded the unresolved spectrum of 2MASSI~J0423485$-$041403 (hereafter 2MASS~J0423$-$0414), because even though it is a known variable (0.8\% $J$-band amplitude,~\citealt{2014ApJ...793...75R}), it is also an astrometric binary system~\citep{2012ApJS..201...19D}.

\begin{deluxetable}{lcccc}
\tablewidth{0pt}
\tablecolumns{5}
\tablenum{1}
\tabletypesize{\scriptsize}
\tablecaption{Confirmed variability sources used as benchmarks in our study.\label{tab:benchmarks}}
\tablehead{
& \multicolumn{2}{c}{Spectral type} & & \colhead{Reference}\\
\cline{2-3}
\colhead{Source} &
\colhead{Optical} & 
\colhead{NIR SpeX} &
\colhead{Young?} &
\colhead{(Optical SpT; Youth; Variability)}
}
\startdata
WISEP J004701.06+680352.1 &  L7pec &  L7 &    Y &  7; 8; 12, 20 \\
SDSSp J010752.33+004156.1 &      L8 &  L7 &    Y &   9;\nodata; 15 \\
SIMP J013656.57+093347.3 &     \nodata &  T2 &    Y & \nodata; 3; 16 \\
2MASSW J0310599+164816 &      L8 &  T0 &    N &   10;\nodata; 1 \\
2MASSI J0825196+211552 &    L7.5 &  L8 &    N &   10;\nodata; 15 \\
SDSS J104335.08+121314.1 &     \nodata &  L9 &    N &   \nodata;\nodata; 15 \\
WISE J104915.57$-$531906.1B &    T1 &  T1 &    N &    11;\nodata; 5 \\
SDSS J105213.51+442255.7 &     \nodata &  T0 &    N &  \nodata;\nodata; 6\\
2MASS J11472421$-$2040204 &     \nodata &  L7 &    Y & \nodata; 18; 15 \\
2MASS J13243553+6358281 &     \nodata &  T3 &    Y & \nodata; 4; 15 \\
2MASS J16291840+0335371 (PSO J247.3273+03.5932) &     \nodata &  T2 &    N &   \nodata;\nodata; 16 \\
2MASS J20025073$-$0521524 &      L6 &  L7 &    N &   2; \nodata;  21 \\
2MASS J21392676+0220226 &     \nodata &  T2 &    Y & \nodata;  23; 17 \\
HN Peg B & \nodata &  T3 &    Y & \nodata; 14; 15\\
2MASS J21481628+4003593 &      L6 &  L7.0 &    Y &   13; 22; 15, 19 \\
\enddata
\tablerefs{(1)~\citet{2015ApJ...798..127B};~(2)~\citet{2007AJ....133..439C};~(3)~\citet{2017ApJ...841L...1G};~(4)~ \citet{2018ApJ...854L..27G};~(5)~\citet{2013AandA...555L...5G};~(6)~\citet{2013ApJ...767...61G};~(7)~\citet{2012AJ....144...94G};~(8)~\citet{2015ApJ...799..203G};~(9)~\citet{2002AJ....123.3409H};~(10)~\citet{2000AJ....120..447K};~(11)~\citet{2013ApJ...770..124K};~(12)~\citet{2016ApJ...829L..32L},~(13)~\citet{2008ApJ...686..528L};~(14)~\citet{2007ApJ...654..570L};~(16)~\citet{2015ApJ...799..154M};(15)~\citet{2014ApJ...797..120R};~(17)~\citet{2014ApJ...793...75R};~(18)~\citet{2016ApJ...822L...1S};~(19)~\citet{2017ApJ...842...78V};~(20)~\citet{2018MNRAS.474.1041V};~(21)~\citet{2019MNRAS.483..480V};~(22)~\citet{2020AJ....160...38V};~(23)~\citet{2021ApJ...911....7Z}}
\end{deluxetable}

\subsection{Selection regions in index-index spaces}

\begin{figure}
    \centering
    \includegraphics[width = 0.9\linewidth]{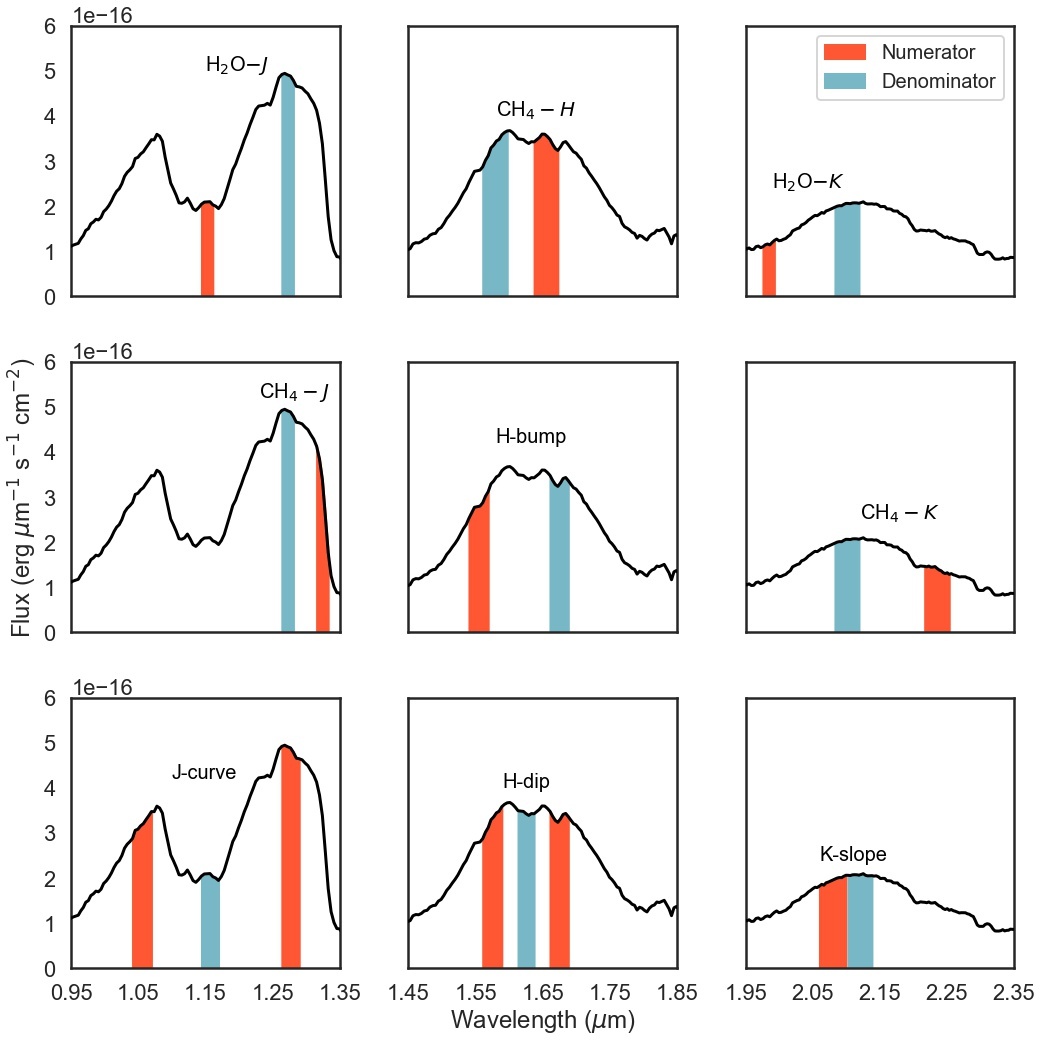}
    \caption{Locations of wavelength regions used to calculate spectral indices superimposed on a sample brown dwarf spectrum. Spectral indices are obtained by integrating the flux column defined by these wavelength regions and finding the ratio between them. For cases where 3 flux columns are used, an average of two flux columns of the same color are used as either numerator or denominator. }
    \label{fig:indices}
\end{figure}

In order to identify signatures of variability from single-epoch spectra, we modified the spectral indices techniques from \citet{2014ApJ...794..143B} and \citet{2010ApJ...710.1142B}, originally designed to identify spectral binary candidates of late-M and L dwarf primaries with a T-dwarf companion. The 13 indices examined are: H$_2$O-$J$, CH$_4$-$J$, $J$-curve, CH$_4$-$H$, H$_2$O-$H$, $K/J$, $J$-slope, H$_2$O-$Y$, $H$-bump, $H$-dip, H$_2$O-$K$, CH$_4$-$K$, and $K$-slope,  which are specified in \citet{2014ApJ...794..143B} and span the NIR range across $J$, $H$, and $K$ bands. The wavelength ranges that correspond to each spectral index are shown in Figure~\ref{fig:indices} and Table~\ref{tab:indices}. We calculated spectral indices by integrating the flux column enclosed by these wavelength regions and finding the ratio between them.

\begin{deluxetable}{lcccc}
\tablewidth{0pt}
\tablecolumns{5}
\tablenum{2}
\tabletypesize{\scriptsize}
\tablecaption{Wavelength Ranges of Spectral Indices\label{tab:indices}}
\tablehead{
\colhead{Spectral Index} &
\colhead{Numerator Range ($\mu$m)} & 
\colhead{Denominator Range ($\mu$m)} &
\colhead{Feature} &
\colhead{Reference}}
\startdata
H$_2$O-$J$     & 1.14-1.165                               & 1.26-1.285                                 & 1.15 $\mu$m H$_2$O          & 1   \\
CH$_4$-$J$     & 1.315-1.335                              & 1.26-1.286                                 & 1.32 $\mu$m CH$_4$          & 1   \\
$J$-curve        & 1.04-1.07 + 1.26-1.29                    & 1.14-1.17                                  & Curvature across $J$-band     & 4   \\
CH$_4$-$H$     & 1.635-1.675                              & 1.56-1.60                                  & 1.65 $\mu$m CH$_4$          & 1   \\
$H$-bump         & 1.54-1.57                                & 1.66-1.69                                  & Slope across $H$-band peak    & 4   \\
$H$-dip          & 1.61-1.64                                & 1.56-1.59 + 1.66-1.69                      & 1.63 $\mu$m FeH/CH$_4$      & 2   \\
H$_2$O-$K$     & 1.975-1.995                              & 2.08-2.10                                  & 1.90 $\mu$m H$_2$O          & 1   \\
CH$_4$-$K$     & 2.215-2.255                              & 2.08-2.12                                  & 2.20 $\mu$m CH$_4$          & 1   \\
$K$-slope        & 2.06-2.10                                & 2.10-2.14                                  & $K$-band shape/CIA H$_2$      & 3  \\
\enddata
\tablecomments{Numerical values of the wavelength regions  of the spectral indices shown in Figure 2. For the indices where the range has a plus sign, the range is the average of the two wavelength ranges listed.}
\tablerefs{(1)~\citet{2006ApJ...637.1067B};~(2)~\citet{2010ApJ...710.1142B};~(3)~\citet{2002ApJ...564..421B};~(4)~\citet{2014ApJ...794..143B}.}
\end{deluxetable}

We measured these indices on our entire spectral sample and compared them against each other to identify regions of variability based on visual inspection. We visually examined 78 index-index plots using \nbenchsp~spectra of \nbench~sources with confirmed variability from photometric monitoring as our benchmarks (see Section~\ref{sec:bench}) to define variability regions in each index-index plot. From these index-index plots, we selected 11 parameter spaces where the benchmark variables were differentiated from the general spectral sample (Figure~\ref{fig:indexindex}). We defined polygonal regions within these plots which conservatively contain all the benchmark variables. The limits for each region in the index-index plots are shown in Table~\ref{tab:regions} and illustrated in Figure~\ref{fig:indexindex}.

We defined variability candidacy based on the number of times a spectrum was found within our variability regions. Figure~\ref{fig:resultshist} shows the frequency with which spectra were identified in a given number of index-index plots. All of our benchmark sources, which are confirmed variables, are found in all 11 index-index plots by design, except for one of the two spectra of the L5.5 dwarf 2MASS J20025073-0521524~\citep[hereafter 2MASS J2002-0521, ][]{2019MNRAS.483..480V}. For this reason, we define weak candidates as sources which fall in 10 regions and strong candidates those which are found in all 11 regions in our index-index plots. This criterion results in \nstrong~strong and \nweak~weak candidate variable sources amenable for future photometric monitoring.

\begin{figure*}[p]
\centering
    \includegraphics[width=0.455\linewidth]{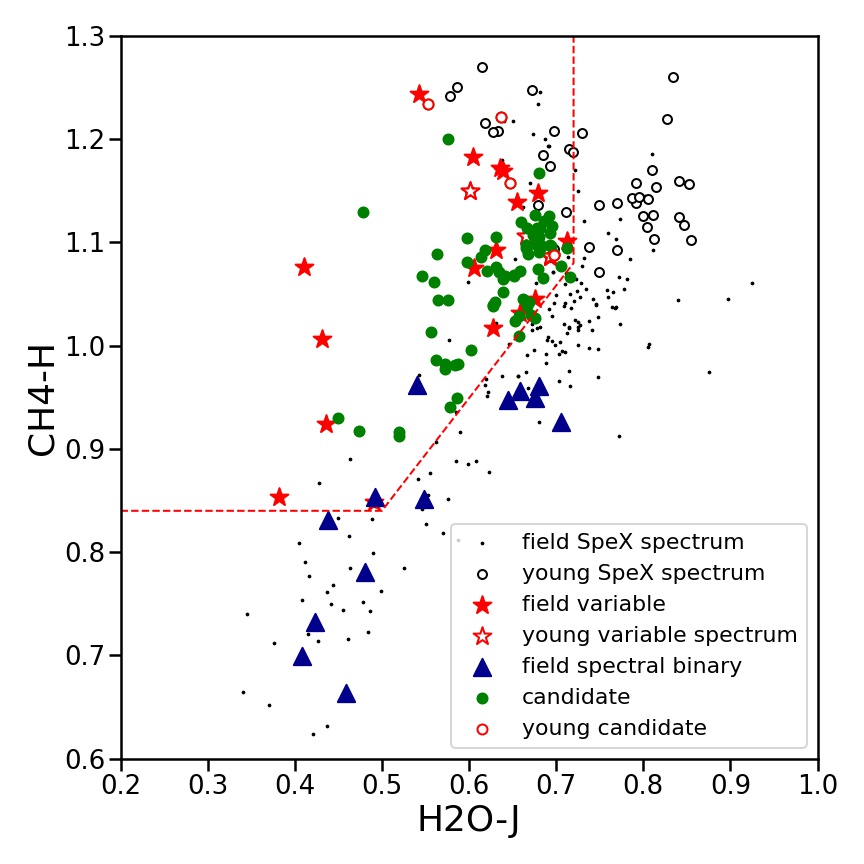}\hfil
    \includegraphics[width=0.455\linewidth]{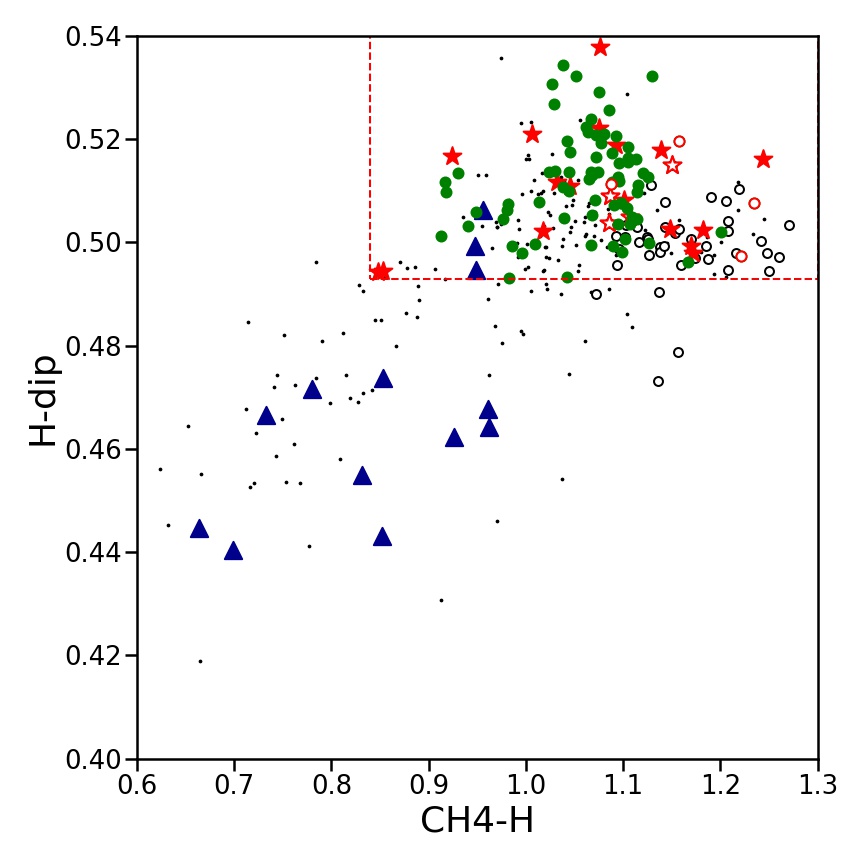}\par\medskip
    \includegraphics[width=0.455\linewidth]{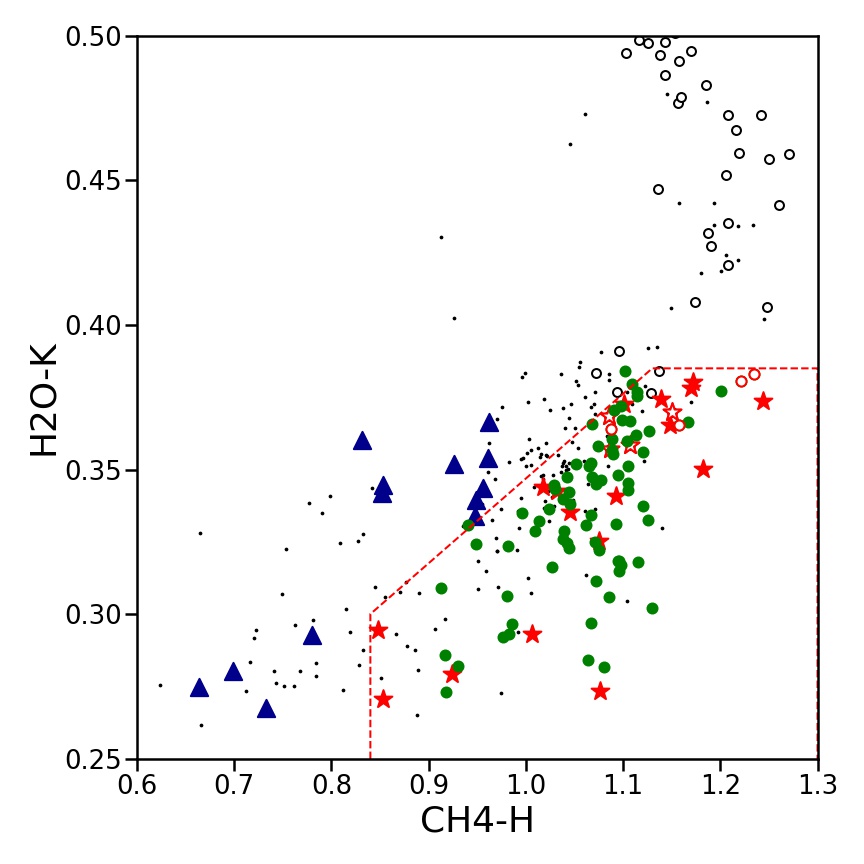}\hfil
    \includegraphics[width=0.455\linewidth]{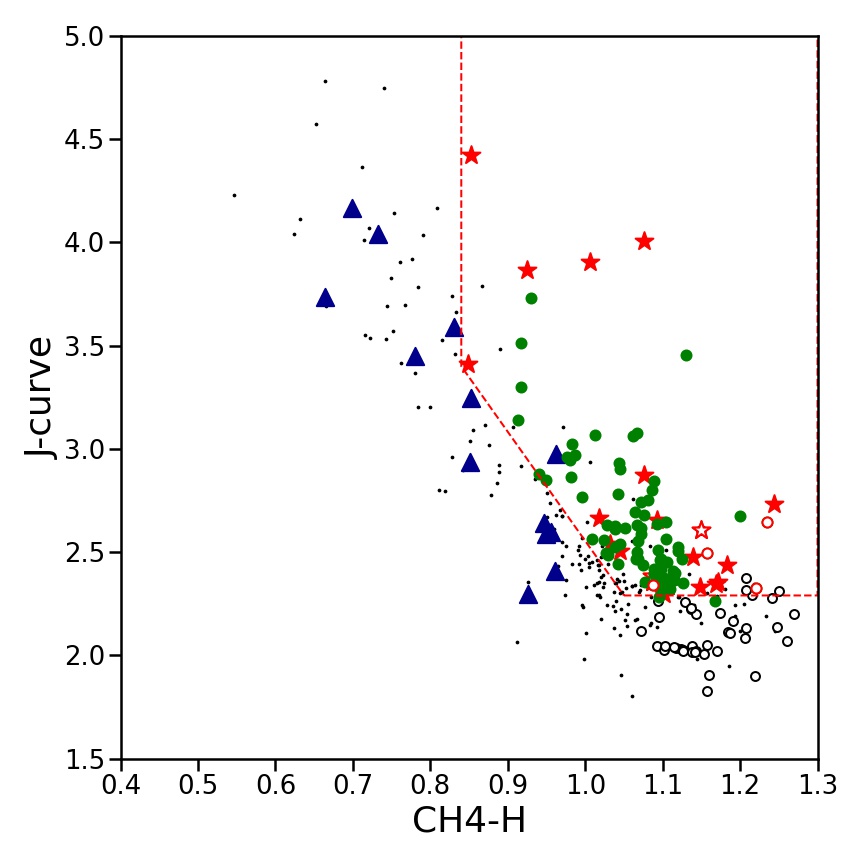}\hfil
    \includegraphics[width=0.455\linewidth]{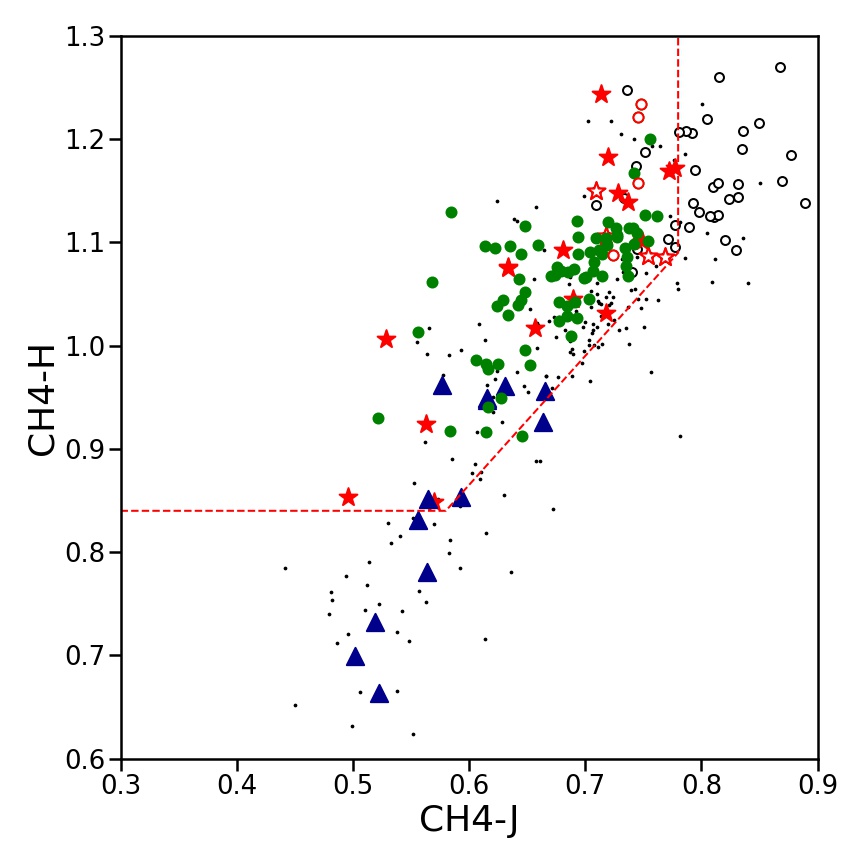}\hfil
    \includegraphics[width=0.455\linewidth]{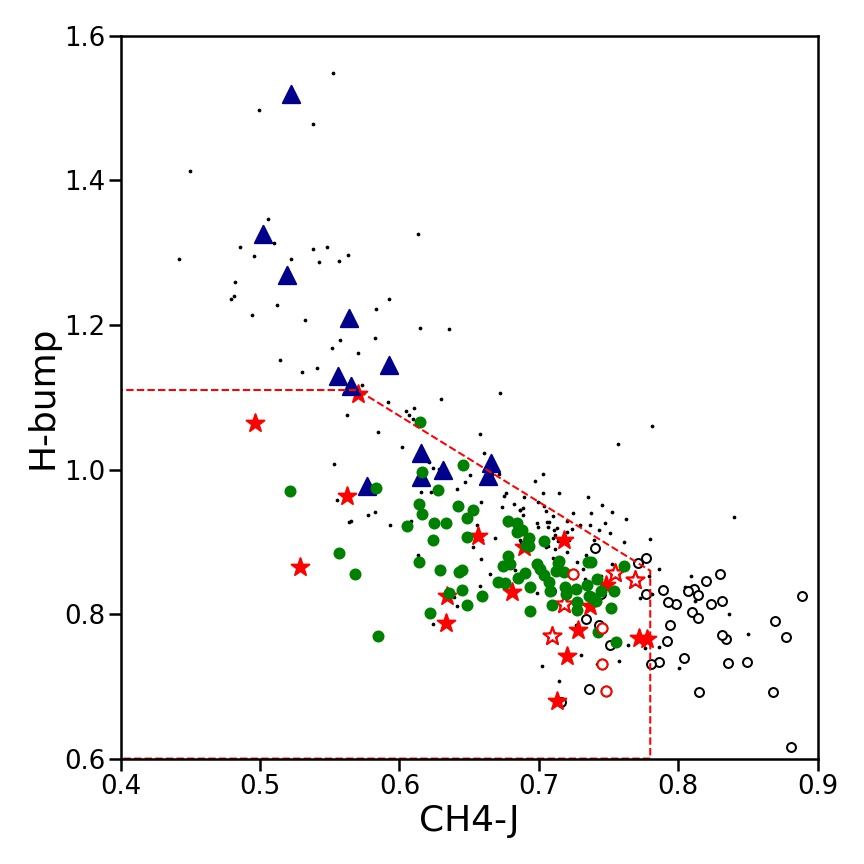}\hfil
\caption{The 11 index-index plots used to identify variability candidates. The red stars indicate our benchmark variables. The blue triangles indicate confirmed spectral binaries. The open stars and open circles indicate youth, while everything that is filled in is a field object.\label{fig:indexindex}}
\end{figure*}

\begin{figure*}[p]
\figurenum{3}
\centering
    \includegraphics[width=0.47\linewidth]{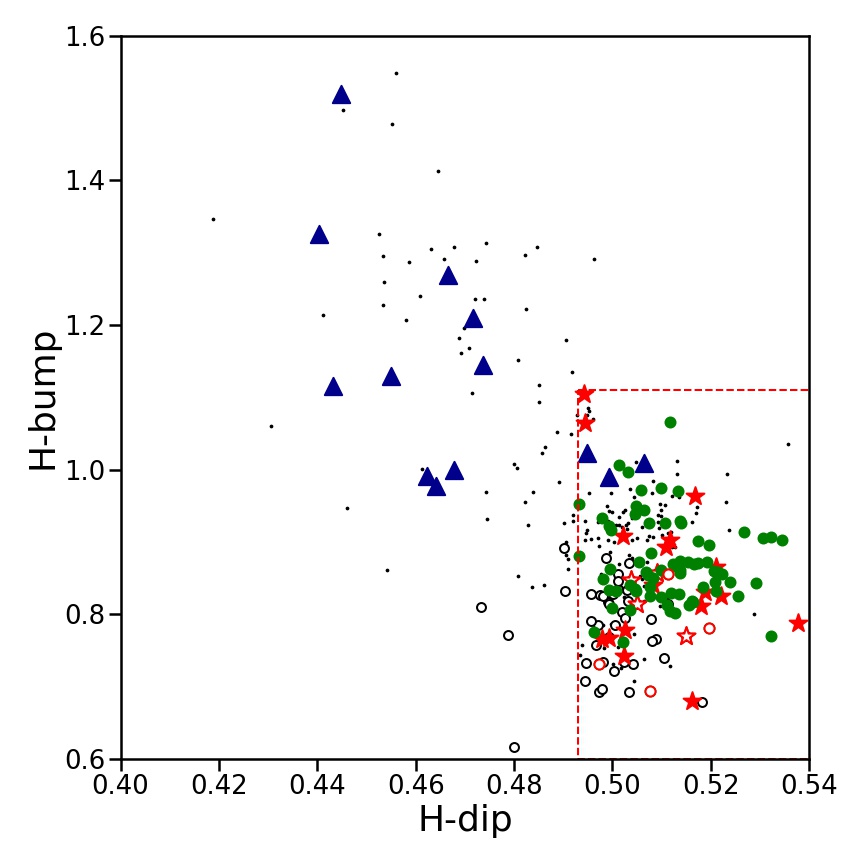}\hfil
    \includegraphics[width=0.47\linewidth]{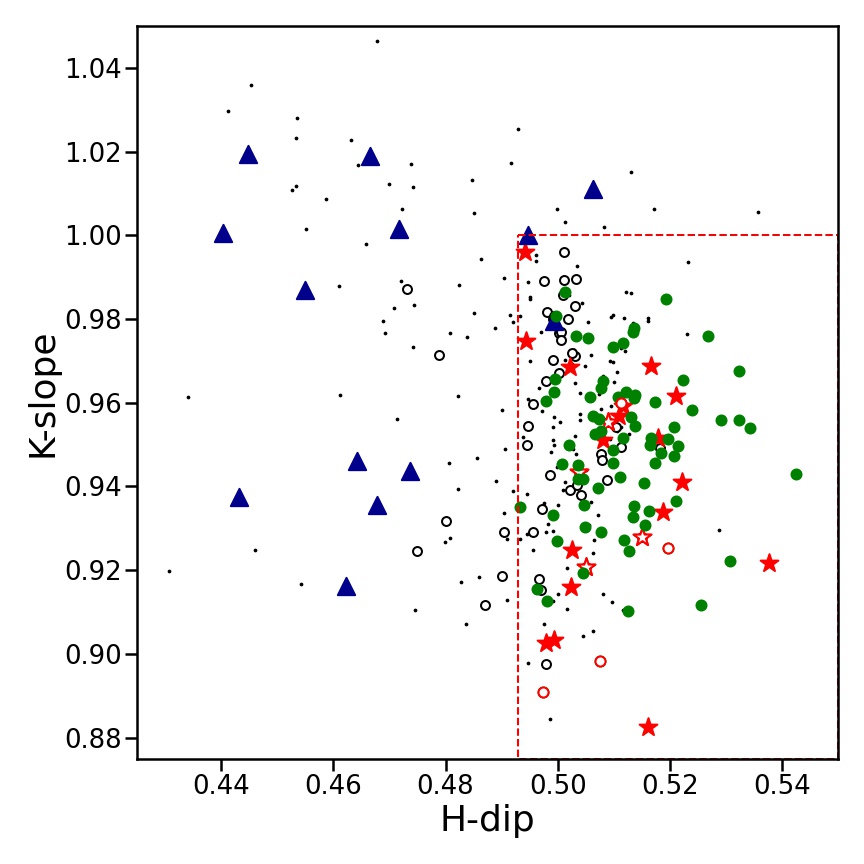}
    \includegraphics[width=0.47\linewidth]{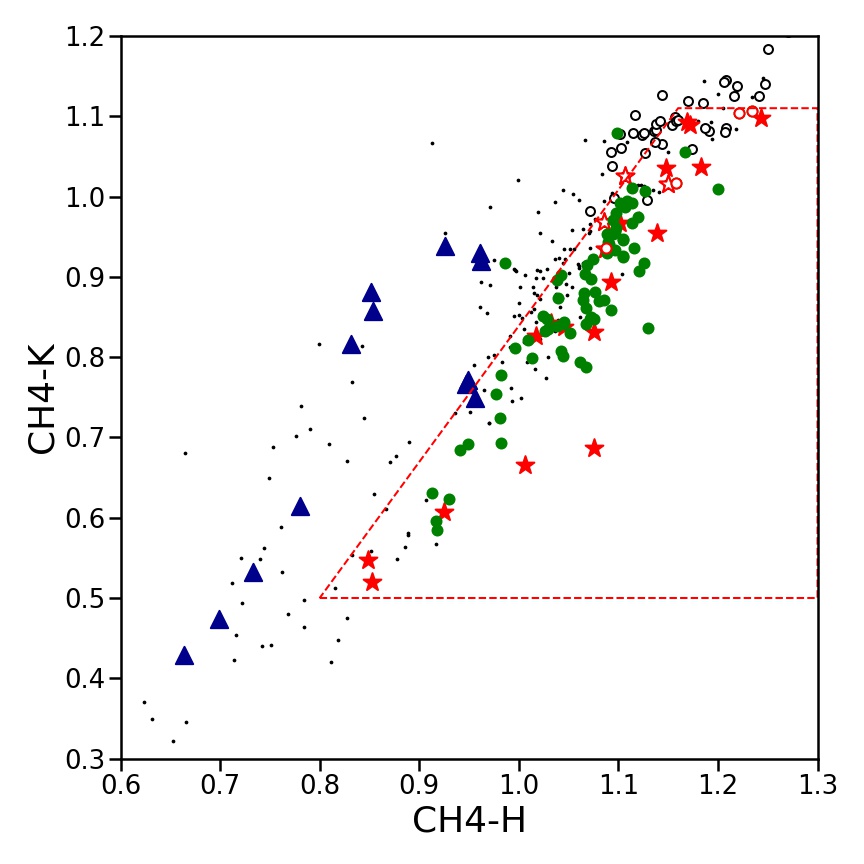}\hfil
    \includegraphics[width=0.47\linewidth]{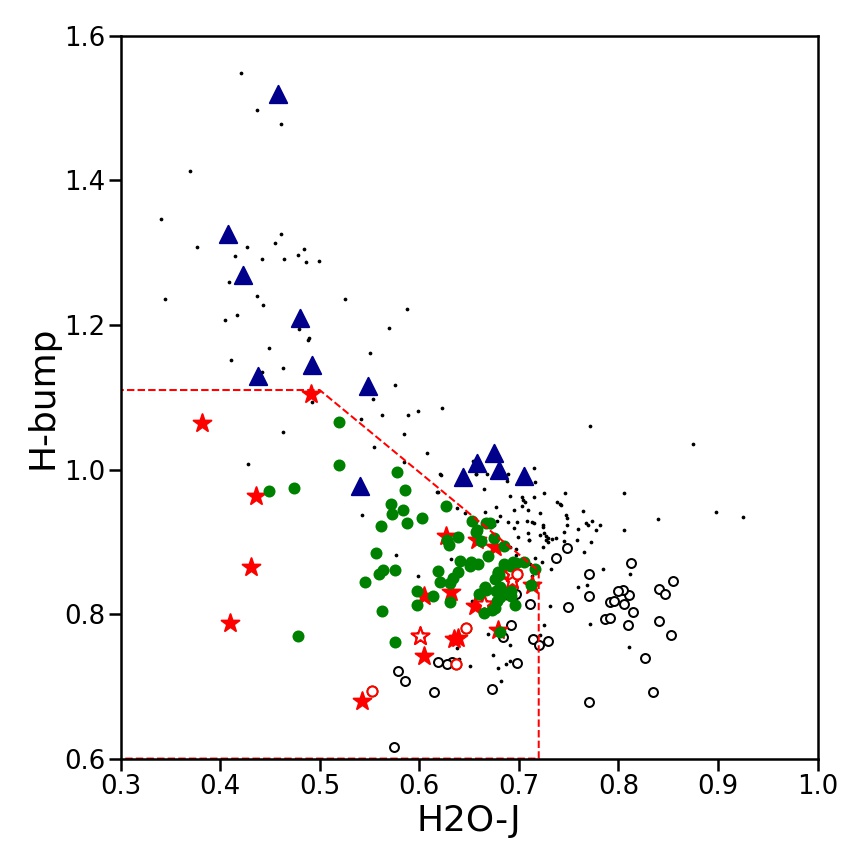}\hfil
    \includegraphics[width=0.47\linewidth]{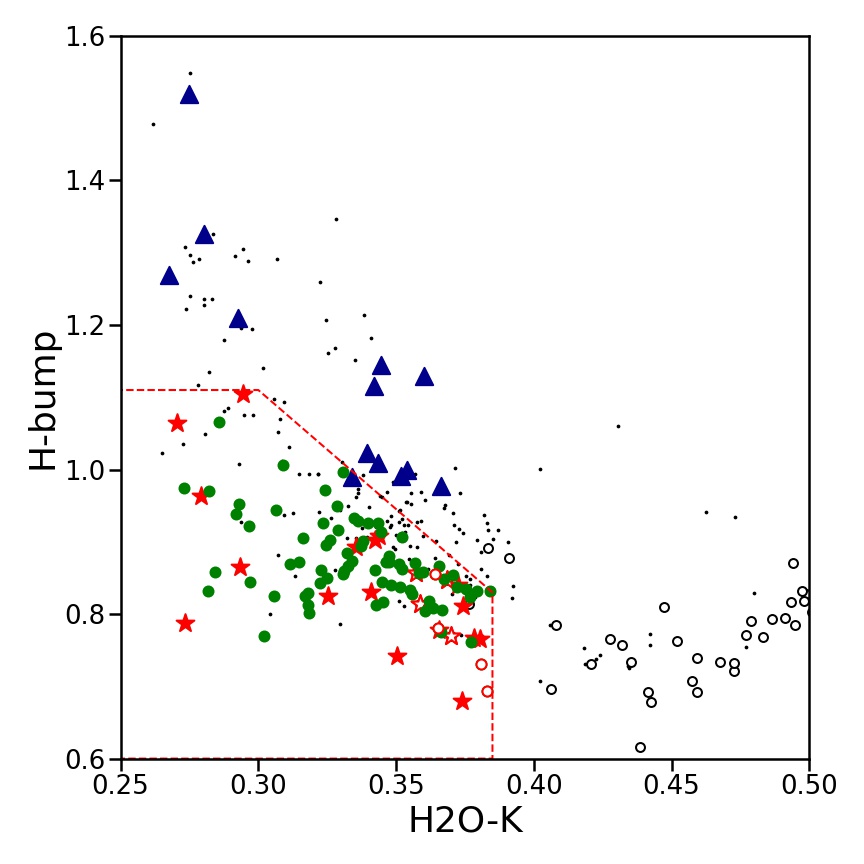}\hfil
\caption{continued}
\end{figure*}

\begin{figure}
    \centering
    \includegraphics[width=0.95\linewidth]{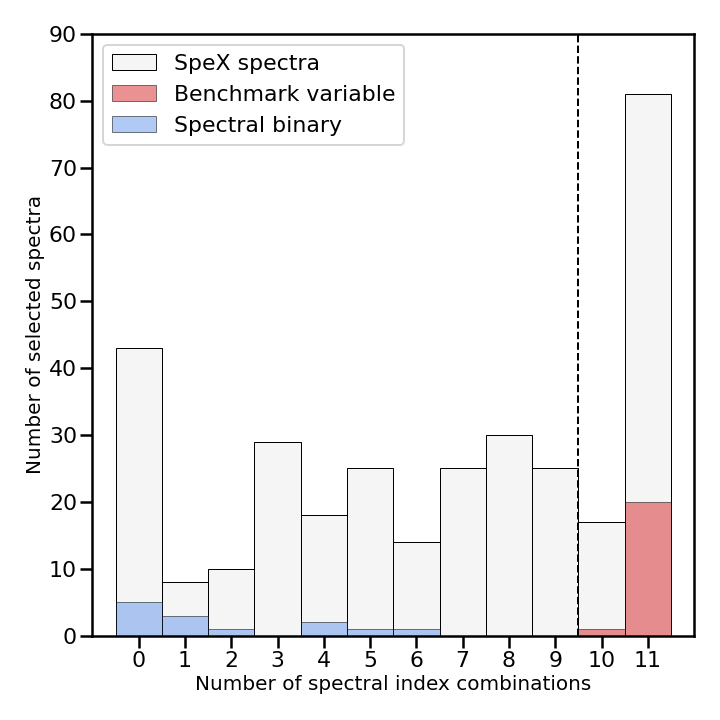}
    \caption{Histogram showing the number of times a spectrum falls in one of our selection regions. Benchmark variables are shown in red, confirmed spectral binaries in blue, and the rest of the spectral sample in gray. The final candidates are the sources found in 10 or more parameter spaces, as shown by the dashed line.}
    \label{fig:resultshist}
\end{figure}

\begin{deluxetable}{lclcc}
\tablewidth{0pt}
\tablecolumns{5}
\tablenum{3}
\tabletypesize{\scriptsize}
\tablecaption{Region limits for each index-index correlation plot.\label{tab:regions}}
\tablehead{
\colhead{x-axis} &
\colhead{y-axis} & 
\colhead{coordinates} &
\colhead{x-range} &
\colhead{y-range}}
\startdata
 H2O-$J$ &    CH4-$H$ &  [[0.2,0.84],[0.5,0.84],[0.72,1.08],[0.72,1.3],[0.2,1.3],[0.2,0.84]] &       [0.2,1] &     [0.6,1.3] \\
 H2O-$J$ &   $H$-bump &  [[0.2,1.11],[0.5,1.11],[0.72,0.86],[0.72,0.6],[0.2,0.6],[0.2,1.11]] &       [0.3,1] &     [0.6,1.6] \\
 CH4-$J$ &    CH4-$H$ &  [[0.3,0.84],[0.58,0.84],[0.78,1.09],[0.78,1.3],[0.3,1.3],[0.3,0.84]] &     [0.3,0.9] &     [0.6,1.3] \\
 CH4-$J$ &   $H$-bump &  [[0.3,1.11],[0.57,1.11],[0.78,0.86],[0.78,0.6],[0.3,0.6],[0.3,1.11]] &     [0.4,0.9] &     [0.6,1.6] \\
 CH4-$H$ &    H2O-$K$ &  [[0.84,0.2],[0.84,0.3],[1.13,0.385],[1.3,0.385],[1.3,0.2],[0.84,0.2]] &     [0.6,1.3] &    [0.25,0.5] \\
 CH4-$H$ &    CH4-$K$ &  [[0.8,0.5],[1.3,0.5],[1.3,1.11],[1.16,1.11],[0.8,0.5]] &     [0.6,1.3] &     [0.3,1.2] \\
 CH4-$H$ &    $H$-dip &  [[0.84,0.493],[1.3,0.493],[1.3,0.54],[0.84,0.54],[0.84,0.493]] &     [0.6,1.3] &    [0.4,0.54] \\
 CH4-$H$ &  $J$-curve &  [[1.05,2.29],[1.3,2.29],[1.3,5.5],[0.84,5.5],[0.84,3.4],[1.05,2.29]] &     [0.4,1.3] &     [1.5,5.0] \\
 H2O-$K$ &   $H$-bump &  [[0.2,1.11],[0.3,1.11],[0.385,0.83],[0.385,0.6],[0.2,0.6],[0.2,1.11]] &    [0.25,0.5] &     [0.6,1.6] \\
 $H$-dip &  $K$-slope &  [[0.493,1],[0.55,1],[0.55,0.875],[0.493,0.875],[0.493,1]] &  [0.425,0.55] &  [0.875,1.05] \\
 $H$-dip &   $H$-bump &  [[0.493,1.11],[0.55,1.11],[0.55,0.6],[0.493,0.6],[0.493,1.11]] &    [0.4,0.54] &     [0.6,1.6] \\
\enddata
\tablecomments{Each set of coordinates forms a polygon with 4 or 5 sides.}
\end{deluxetable}

\subsection{Notable Sources}

\subsubsection{Strong Candidates}

\textbf{1) WISE J00164397+2304265}

This object was first identified following a crossmatch of the PanSTARRS \citep{2010SPIE.7733E..0EK} and WISE \citep{2010AJ....140.1868W} surveys as a T0 brown dwarf by \citet{2015ApJ...814..118B}. The $H$-band showed no signs of methane, which is a hallmark of the T dwarf class~\citep{1999ApJ...522L..65B}, but the $J$-band peak resembles that of a T2 dwarf. Thus it was flagged as a spectral binary candidate for its unusual spectrum \citep{2015ApJ...814..118B}, finding a closest match with the confirmed L6 + T2 dwarf binary, SDSS J0423–0414 \citep{2005ApJ...634L.177B,2002ApJ...564..466G}. This source has a 93\% probability of belonging to the field using parallaxes from \citet{2020AJ....159..257B} in BANYAN $\Sigma$~\citep{2018ApJ...856...23G}.


\medskip

\textbf{2) 2MASS J00310928+5749364}

2MASS J0031+5749 was classified as an L9 brown dwarf by \citet{2013ApJ...777...84B} using spectral indices from \citet{2006ApJ...637.1067B}. This source has a 88\% membership  probability in Carina Near according to BANYAN $\Sigma$~\citep{2018ApJ...856...23G} with kinematic measurements from~\citet{2015ApJ...814..118B}, giving it an age estimate of $13.3^{+1.1}_{-0.6}$ Myr \citep{2021MNRAS.500.5552B}. This source was independently monitored for variability with the \emph{Spitzer Space Telescope} and found to be variable by ~\citet{2022ApJ...924...68V}, who measured an amplitude of 0.35$\pm$0.03\% at [3.6] and a period of 1.64$\pm$0.01\,h. The SPL contains two spectra of this object, both found in all 11 index-index regions.

\medskip

\textbf{3) 2MASS J02062493+2640237}

The $J$-band spectrum of this object matches that of an L9 dwarf, but the $H$ and $K$-bands are unusually red, such that it was classified as an L9pec by~\citet{2011ApJS..197...19K}. This object is a part of a rare class of L dwarfs that are red for reasons not attributable to low gravity and appear to have older kinematics than field L dwarfs \citep{2011ApJS..197...19K}. This object has a 66.9\% probability of belonging to the field according to BANYAN $\Sigma$. 

\medskip

\textbf{4) WISE J023038.90-022554.0}

This object was classified as a peculiar L8 dwarf by \citet{2013PASP..125..809T} using spectral standards from \citet{2010ApJS..190..100K}, who noted the object's $H$-band divot near 1.6\,$\mu$m as well as methane at 2.2\,$\mu$m, which are typically indicative of an unresolved T dwarf companion. \citet{2013PASP..125..809T} suggested that this spectrum could better be explained as a spectral binary of L7 and T2 components instead of a singular L8 object, however no high resolution imaging exists of this object. According to BANYAN $\Sigma$~\citep{2018ApJ...856...23G}, this object has a 48.6\% membership probability in the Hyades (an age of $625 \pm 0.5$ Myr; \citet{2015ApJ...807...58B}) and 34.3\% in Carina Near ($13.3^{+1.1}_{-0.6}$ Myr). 

\medskip



\textbf{5) 2MASS J03264225-2102057}

The NIR spectrum of this object was noted by~\citet{2007AJ....133..439C} to be redder than typical for its spectral type and therefore younger than 500\,Myr.~\citet{2015ApJS..219...33G} later classified 2MASS J0326-2102 as an L5$\beta$, the $\beta$ signifying intermediate surface gravity. This object is also a high likelihood member of the AB Doradus moving group ($130 \pm 20$ Myr; \citet{2004ApJ...614..386B}). 0326-2102 shows a variability amplitude of 0.34\% $\pm$ 0.04, albeit a long period of $32.84^{+4.69}_{-4.62}$\,h, in the \emph{Spitzer} survey of~\citet{2022ApJ...924...68V}.

\medskip

\textbf{6) 2MASS J06420559+4101599}

This object is an unusually red L9 dwarf. Using solely the $J$-band, this object would result in a T spectral type but shows no sign of CH$_{4}$, so the redness could result from a dusty photosphere. It was classified as a medium binary candidate by \citet{2015ApJ...814..118B} using criteria from \citet{2010ApJ...710.1142B}. This object was given a 78.6\% probability of membership in the AB Doradus Moving Group by the BANYAN II online tool \citep{2015ApJ...798...73G}, yet updated with BANYAN $\Sigma$ to 86\% membership probability in AB Doradus \citep{2022ApJ...924...68V} and thus an age of approximately $130 \pm 20$ Myr. This source has been confirmed as a photometric variable in the \emph{Spitzer} survey of ~\citet{2022ApJ...924...68V} with an amplitude of $2.16 \pm 0.16$\% in Spitzer and period of $10.11 \pm 0.06$\,h.

\medskip

\textbf{7) 2MASS J06522224+4127366 (PSO J103.0927+41.4601)}

Using spectral standards from \citet{2006ApJ...637.1067B}, this object was classified as a T0.5. However, \citet{2013ApJ...777...84B} discovered that this object is a close spectral match to the L6+T2 binary SDSS J0423-0414AB, one of the known variable objects for which only the blended light spectrum was available in our spectral library. This is a field source based on BANYAN $\Sigma$.

\medskip

\textbf{8) SDSS J075840.33+324723.4}

This source was selected as a weak spectral binary candidate of T0.0+T3.5 dwarf components by~\citet{2010ApJ...710.1142B}. It is classified as a T2 dwarf from its SpeX spectrum~\citep{2004AJ....127.3553K}, being only slightly bluer in the $J$-band than its best-fit single template. \emph{HST} imaging did not resolve this source into two components~\citep{2006ApJS..166..585B}. SDSS J0758+3247 has a 93.9\% membership probability in Carina Near ($13.3^{+1.1}_{-0.6}$ Myr) according to BANYAN $\Sigma$~\citep{2018ApJ...856...23G} and a $4.8\pm0.2\%$ variability amplitude in $J$-band from a Spitzer Space Telescope variability monitoring campaign \citep{2020AJ....160...38V}.

\medskip

\textbf{9) SDSS J080959.01+443422.2}

\citet{2006AJ....131.2722C} identified this object as an L6 spectral type brown dwarf from \citet{2010ApJ...710.1142B} standards. \citet{2016ApJS..225...10F} proposed that this object is an ambiguous member of either the Argus (40 - 50 Myr; \citet{2019ApJ...870...27Z}) or AB Doradus ($130 \pm 20$ Myr) moving groups from its spectrum and kinematics. An ambiguous member requires updated astrometric precision as it either appeared to be part of more than one group or could not be differentiated from the field. This object was observed twice with SpeX, with one of its spectra found in all 11 of our 11 index-index regions, while the other one is only found in 10. This source shows a flat light curve in the \emph{Spitzer} survey of ~\citet{2022ApJ...924...68V}.
\medskip



\medskip

\textbf{10) 2MASS J08300825+4828482}

2MASS J0830+4828 was classified as an L9.5 object by \citet{2014AJ....147...34S} using the L dwarf classification scheme in \citet{2010ApJS..190..100K}. H$\alpha$ emission was detected in this object by \citet{2007AJ....133.2258S}, indicative of magnetic activity and suggestive of youth. However, according to BANYAN $\Sigma$, this object has a 99.9\% probability of having field kinematics. The SPL contains two spectra of this object which were both found in all 11 index-index regions.
\medskip

\textbf{11) SDSS J083717.21-000018.0}

This object was one of the earliest L/T transition objects discovered \citep{2000ApJ...531L..57B} and classified as a T1 brown dwarf. Of note, it is one of a handful of T-dwarfs with H$\alpha$ emission \citep{2008ApJ...689.1295K}. Its kinematics are consistent with field membership (99.9\% probability). This object has 3 spectra in our library, two appearing in all 11 index-index regions, and one in only 10.
\medskip

\textbf{12) 2MASS J08503593+1057156}

 \citet{2011AJ....141...70B} posited that this might be a triple system with an equal mass unresolved binary due to the fact that one of the objects was twice as bright as the other despite being of the same spectral type. Later, \citet{2011AJ....141...71F} identified a potential binary comover. The SpeX spectrum of 2MASS J0850+1057 was later modeled as a L6.5$\pm$1 and L8.5$\pm$1 spectral binary and the comover was disproved with more precise astrometry \citep{2012ApJS..201...19D}. However, HST photometry for this object \citep{2021ApJ...914..124B}  suggested the large differences in component brightness across several bands hinted that the A component is indeed a binary itself, as originally suggested by \citet{2011AJ....141...70B}. This object was also originally thought to have signatures of youth, but \citet{2015ApJS..219...33G} concluded that it is most likely a field object, positing that its unusually red spectrum could be explained by the presence of thicker/higher clouds in the atmosphere. There are 2 spectra of this system in our SpeX Prism Library, both of them appearing in all 11 index-index regions.
\medskip

\textbf{13) SDSS J085234.90+472035.0}

Using the L dwarf classification scheme developed by \citet{2010ApJS..190..100K}, this object was classified as an L9.5 dwarf \citep{2014AJ....147...34S}. While thought to  potentially belong to a young moving group \citep{2015ApJS..219...33G}, with the updated BANYAN $\Sigma$ it has  99.9\% probability of being part of the field. \citet{2016MNRAS.455.1341M} identified SDSS J0852+4720 as a weak candidate for spectral binarity of L7+T1 dwarf components using the spectral indices from \citet{2014ApJ...794..143B} and \citet{2010ApJ...710.1142B}.
\medskip



\medskip

\textbf{14) SDSS J093109.56+032732.5}

SDSS J0931+0327 is unusually blue for its spectral type of L7.5 \citep{2004AJ....127.3553K}. According to BANYAN $\Sigma$, this object has a 89.9\% probability of belonging to the Carina Near moving group ($13.3^{+1.1}_{-0.6}$ Myr).  Using cloud models, \citet{2004AJ....127.3553K} calculated that its unusually blue colors might indicate a high sedimentation efficiency and a low cloud optical depth. \citet{2014ApJ...793...75R} obtained a light curve for this object as a part of a variability survey and did not detect significant variability. They detected a peak to peak variability amplitude of 0.041\%, which was not significant as the white noise estimate for the relative-flux light curve was 0.036\%. This object was identified as a spectral binary candidate by \citet{2015AJ....150..163B} through visual inspection, yet an adaptive optics image with Keck/NIRC2 did not reveal a secondary companion~\citep{2015AJ....150..163B}. 

\medskip

\textbf{15) 2MASS J09553336-0208403}

2MASS J0955-0208 is a known young L7 brown dwarf \citep{2017ApJS..228...18G}, although its membership probabilities point to 55.5\% for field and 33.5\% for AB Doradus ($130 \pm 20$ Myr). This object has a very red $J-K$ color, a triangular-shaped $H$-band continuum, and weak K I absorption lines, which are all indicative of low surface gravity.  
\medskip

\textbf{16) 2MASS J09560810-1447065}

 \citet{2015ApJ...814..118B} classified this brown dwarf as an L9 type using visual comparisons with field standards and a T0.0$\pm$0.5 type brown dwarf using index-based classification. They flagged this object as a medium binary candidate because of subtle signs of methane absorption indicative of a T-dwarf using the spectral indices method \citep{2010ApJ...710.1142B,2014ApJ...794..143B}. 2MASS J0956-1447 is a known field object \citep{2017AJ....153..196S}. 

\medskip

\textbf{17) 2MASSI J1043075+222523}

\citet{2007AJ....133..439C} discovered this object and performed an optical spectroscopy follow-up of this object and classified it as an unusually red L8 dwarf. They provided the possible explanation that the object was an unresolved spectral binary as a reason for its peculiar spectrum. 2MASS J1043+2225 is classified as an intermediate gravity object with the scheme from \citet{2013ApJ...772...79A}. This object has measured H$\alpha$ emission~\citep{2016ApJ...826...73P} and a measured rotation period of $2.21^{+0.14}_{-0.13}$ hours in 8-12\,GHz radio frequencies~\citep{2018ApJS..237...25K}.

\medskip

\textbf{18) Luhman 16A (WISE J104915.57$-$531906.1A)}

Luhman 16A was discovered in 2013 by \citet{2013ApJ...767L...1L} using multi-epoch astrometry from the Wide-field Infrared Survey Explorer (WISE), as one component of a 3\,AU separated flux reversal binary, meaning that  the two components switch in relative brightness at different bands. The B component is brighter at $J$, but fainter
in the $K$ band, which is a phenomenon that can potentially be explained by thick clouds \citep{2013ApJ...772..129B}. Luhman16AB is located at $1.9980\pm0.0004$\,pc, making it the third closest system to the Sun after Alpha Centauri ($d = 1.3475\pm0.256025$\,pc;~\citealt{2018AJ....155..265H}) and Barnard's star ($d = 1.8282\pm0.00013$\,pc;~\citealt{2021A&A...649A...2L}). \citet{2013ApJ...772..129B} classified this object as an L7.5$\pm$0.9 spectral type brown dwarf, and its companion as a T0.5$\pm$0.7 dwarf using indices and spectral type/index relations \citep{2002ApJ...564..466G, 2006ApJ...637.1067B,2007ApJ...659..655B}. Luhman 16A is one of the reddest L dwarfs known ($J-K = 2.08\pm0.08$\,mag,~\citealt{2013ApJ...772..129B}), suggestive of thick cloud coverage \citep{2014ApJ...790...90F}.

The individual components of Luhman 16 have been monitored for variability multiple times in the past. \citet{2015ApJ...798..127B} used \emph{HST}/WFC3 at $1.1$–$1.66\,\mu$m for spectroscopic variability and found an upper limit for Luhman 16A of 0.4\%, whereas the B component showed variability amplitudes of 7-11\% changing within a single rotation period. Later \emph{HST}/WFC3 observations in the $0.8$-$1.15\,\mu$m range found a 4.5\% amplitude in Luhman 16A \citep{2015ApJ...812..163B}. Using TESS data of the system, the period of Luhman 16A was determined to be 6.94 hours, and this periodicity is much lower amplitude than that of the B component \citep{2021ApJ...906...64A}. Mutli-wavelength photometric monitoring with the 2.2-m MPG/ESO GROND telescope covering $rizJHK$ bands found 2\% and 3\% variability amplitudes for Luhamn 16A in i and z bands, respectively \citep{2013ApJ...778L..10B}. However, no variability was detected in r-band or the NIR $JHK$ bands. The NIR spectrum of Luhman 16A can be modeled with an intermediate thickness, single cloud layer at 1200\,K~\citep{2015ApJ...798..127B}. The B component of this binary system is a known variable brown dwarf and a benchmark variable in our analysis. 


\medskip

\textbf{19) 2MASS J11193254-1137466}

2MASS J1119-1137 or TWA 42 is one of the reddest sources known
with a $J-K$ color of $2.62 \pm 0.15$\,mag \citep{2015AJ....150..182K}. This source is a high likelihood member of the 10\,Myr old TW Hydrae association according to its kinematics and spectra \citep{2016ApJS..225...10F}. However, updated kinematics yield a BANYAN Sigma estimate of 63.2\% probability of field membership \citep{2020AJ....159..257B}. TWA 42 was discovered to be an approximately equal magnitude binary of L7 dwarfs with a separation of $0\farcs14$ ($3.6 \pm 0.9$\,AU;~\citealt{2017ApJ...837...95B}). Both components are among the lowest-mass brown dwarfs in the solar neighborhood ($M\approx4 M_\mathrm{Jup}$). \citet{2018AJ....155..238S} performed time-series photometry of this object with \emph{Spitzer} [3.6] and [4.5] bands and found low amplitude variability of $0.230^{+0.036}_{-0.0035}\%$ and  $0.453\pm0.037\%$ , respectively, but could not quantify the effects of binarity on the system's rotational parameters.

\medskip

\textbf{20) SDSS J121951.45+312849.4}

SDSS J1219+3128 is an L9 dwarf \citep{2014AJ....147...34S}, and known variable from the \textit{Hubble Space Telescope} Near-Infrared Spectroscopic Survey for variability, but it only has a lower limit for variability ($>2\%$) as they were not able to observe the full period \citep{2014ApJ...782...77B}. Therefore we did not include it in our benchmark variable objects. \citet{2019AJ....157..101M} searched for whether confirmed photometrically variable objects also were candidates for having composite spectra using the \citet{2010ApJ...710.1142B} and \citet{2014ApJ...794..143B} techniques and found that it was a strong spectral binary candidate with a best matching fit of an L8 and T4 ($\Delta T_\mathrm{eff} \approx$200\,K).
\medskip

\textbf{21) SDSSp J125453.90-012247.4}

SDSSp J1254-01222 is a T2 dwarf 
with weak $H\alpha$ emission at $6563$\,\AA \citep{2003ApJ...594..510B}. \citet{2015ApJ...799..154M} detected an upper limit of variability for this object of $<0.15\%$ at [3.6] and $<0.3$\% at [4.5] with \emph{Spitzer}, which is very low-level variability. This object is also a field age dwarf \citep{2015ApJ...810..158F}.
 
 \medskip

\textbf{22) DENIS J132620.0-272936}

DENIS J1326-2729 is an L5 dwarf in the optical \citep{2002ApJ...575..484G} and an L7 in the NIR. This object has a spectrum sloped towards red wavelengths and a characteristic peaked $H$-band suggesting youth. While this object has been flagged as a potential member of the Argus young moving group \citep{2015ApJS..219...33G} meaning an age estimate of 40 - 50 Myr, BANYAN $\Sigma$ gives an 88\% probability of membership in Carina Near ($13.3^{+1.1}_{-0.6}$ Myr). 
\medskip

\textbf{23) SDSS J133148.92-011651.4}

SDSS J1331-0116 is an L8 brown dwarf in the infrared with unusually blue colors for its spectral type, similar to another candidate SDSS J0931+0327 \citep{2004AJ....127.3553K}, potentially due to low metallicity. Its atypical spectral features usually point to a blend of binary compositions, but \citet{2021AJ....161...42B} calculated a binary spectral decomposition of L6+T5 with $\Delta J = 1.97 \pm 0.18$\,mag which is much larger than the observed magnitude difference of 0.6\,mag, making it unlikely that the peculiarity is due to binarity. This object has three observations in the SPL, which were all found in all 11 index-index regions.



\medskip

\textbf{24) 2MASS J15164306+3053443}

This object is a weak spectral binary candidate with no high resolution imaging \citep{2010ApJ...710.1142B, 2015AJ....150..163B}, potentially composed of an L8+L9.5 dwarfs. However, the binary template fitting of \citet{2010ApJ...710.1142B} could not rule out the possibility that this object was single with an unusual atmosphere. 2MASS J1516+3053 is also a confirmed variable object from \emph{Spitzer} \citep{2015ApJ...799..154M} with amplitudes of $2.4 \pm 0.2\%$ at [3.6] and $3.1 \pm 1.6$\% at [4.5] bands and a period of $6.7$\,hr.


\medskip

\textbf{25) 2MASSW J1632291+190441}

2MASSW J1632+1904 is one of the earliest discovered late L dwarfs \citep[L8,][]{1999ApJ...521..613R}, and identified as a weak binary candidate by \citet{2010ApJ...710.1142B}. 
\citet{2015ApJ...799..154M} found this object to have amplitudes of $0.42 \pm 0.08\%$ at [3.6] and $0.5 \pm 0.3$\% at [4.5] and a period of $3.9 \pm 0.2$\,hr, and thus we did not use it as a benchmark variable due to its low-level variability ($<1\%$). This object contains two spectra in the SPL.
\medskip

\textbf{26) WISE J164715.57+563208.3}

This L9 brown dwarf is unusually red with $J-K = 2.20 \pm 0.10$\,mag. In a red optical survey with Keck/LRIS, \citet{2016ApJ...826...73P} obtains optical spectra for WISE J1647+5632 and suggests that it is an unresolved binary system from its spectra. This object is unlikely to be young from its proper motion, photometric distance estimate, and spectrum \citep{2017AJ....153..196S}. This source has been independently confirmed as variable by the \emph{Spitzer} survey of ~\citet{2022ApJ...924...68V} with an amplitude of $0.47\pm0.06$\% and period of $9.234_{-0.25}^{+0.23}$ hours.

\medskip

\textbf{27) WISE J173859.27+614242.1}

WISE J1738+6142 was discovered in 2013 with WISE. This object is a late L/early T, but it defies a standard spectral classification because of its highly unusual spectrum \citep{2013ApJS..205....6M}. It has excess flux on the blue side of the $H$ and $K$ bands, hinting at a T dwarf secondary and is thought to have rapid changes in atmospheric conditions. 

\medskip

\textbf{28) 2MASS J17410280-4642218}

This object is an L5 dwarf and was noted to likely be a member of $\beta$-Pic or AB Doradus moving groups \citep{2016ApJS..225...10F}. 2MASS J1741-4642 is now a confirmed member of AB Dor \citep{2021ApJS..253....7K} and thus $130 \pm 20$ Myr old. \citet{2019MNRAS.483..480V} observed this object for photometric variability using the New Technology Telescope (NTT) and the United Kingdom Infrared Telescope (UKIRT) using the $J_S$ band (1.16-1.32\,$\mu$m), but did not observe any variability. However, the NTT observation was of short duration and therefore amplitudes would have had to be greater than 10\% to detect significant variations. This source has been confirmed as a photometric variable in the \emph{Spitzer} survey of ~\citet{2022ApJ...924...68V} with an amplitude of $0.35\pm 0.03$ and period of $15_{-0.57}^{+0.71}$ hours.
\medskip

\textbf{29) WISEP J183058.57+454257.9}

This object was first identified by \citet{2011ApJS..197...19K} with WISE and classified as an L9 dwarf. It was identified as a medium binary candidate \citep{2015ApJ...814..118B} using the criteria from \citet{2010ApJ...710.1142B}, showing signs of methane absorption. 
\medskip

\textbf{30) WISE J185101.83+593508.6}

WISE J1851+5935 was classified as L9 when evaluated as a single template \citep{2013ApJ...777...84B}. However, the $H$ and $K$-band peaks have lower flux than typical L9 standards \citep{2010ApJS..190..100K}. \citet{2013PASP..125..809T} analyzed this spectrum as an unresolved binary, and found that it was better fit as an L7 and T2 binary than as a single source. However, WISE J1851+5935 does not appear as a spectral binary candidate in \citet{2010ApJ...710.1142B} or \citet{2014ApJ...794..143B}.
\medskip

\textbf{31) 2MASSW J2101154+175658}

This object was discovered by \citet{2000AJ....120..447K}.
It was identified as a binary system by \citet{2003AJ....126.1526B} and composed of L7 and L8 dwarfs \citep{2012ApJS..201...19D}. 2MASSW J2101+1756 is a potential member of the Scorpius Centaurus Complex (SCC) \citep{2015ApJS..219...33G} which is $\sim 10$ Myr old \citep{2012AJ....144....8S}, but the updated Banyan $\Sigma$ gives a 56\% field probability. There are two spectra of this object in the SPL.

\medskip

\textbf{32) 2MASS J21243864+1849263}

From a crossmatch of SDSS, 2MASS, and WISE, 2MASS J2124+1849 was identified as an L-dwarf with unusual red colors that could suggest low surface gravity or unusual dust content and cloud properties. \citep{2015AJ....150..182K}. According to BANYAN $\Sigma$, this object has a 79.7\% membership probability in Carina Near $13.3^{+1.1}_{-0.6}$ Myr.
\medskip

\textbf{33) 2MASS J21315444-0119374}

This object was identified from imaging data of the Sloan Digital Sky Survey (SDSS) \citep{2006AJ....131.2722C}. \citet{2014AJ....147...34S} classified this object as an L9.5 in the near infrared using the L dwarf classification scheme found in \citet{2010ApJS..190..100K}. This object is a widely separated common proper motion companion of the low-mass stars NLTT 51469AB, which is a close binary of M3 and M6 components \citep{2019MNRAS.487.1149G}. 
\medskip

\textbf{34) 2MASS J22153705+2110554}

2MASS J2215+2110 was discovered very recently by \citet{2015AJ....150..182K}, as it was previously overlooked in SDSS. This T0.0 dwarf is only a weak candidate binary using the indices from \citet{2010ApJ...710.1142B}, but it has a peculiar spectrum and a lack of flux in the $H$-band compared to other early T dwarfs. \citet{2019AandA...629A.145E} identified this object as a strong variable with a $10.7\pm0.4\%$ amplitude in the $J$ band and noted that it seems unlikely that this object is an unresolved binary system as previously suggested. We did not include this object as one of our benchmark objects, because it was discovered to be variable after the curation of our benchmark list. However, the SPL contains two observations of this object both found in all 11 index-index regions.
\medskip

\textbf{35) 2MASS J22443167+2043433}

This object is an L6 spectral type brown dwarf \citep{2016ApJS..225...10F}. 2MASS~J2244+2043 is a bona fide member of AB Doradus \citep{2016ApJS..225...10F} meaning that it is $130 \pm 20$ Myr old. \citet{2006ApJ...653.1454M} detected sinusoidal variability with a period of 4.6 hours for this object using the Spitzer/IRAC 4.5\,$\mu$m and 8\,$\mu$m band passes. \citet{2018MNRAS.474.1041V} identified a longer period of 11$\pm$2\,h with \emph{Spitzer} light curves, seeing a variability amplitude of 0.8$\pm$0.2\% in [3.6]. However, \citet{2019MNRAS.483..480V} observed a 5.5$\pm$0.6\% variability amplitude in $J$-band. The change in period is most likely due to the short observing window of 8.2\,h in \citet{2006ApJ...653.1454M}.
\medskip

\textbf{36) WISE J232728.74-273056.6}

WISE J2327-2730 is an L9 object with slightly redder WISE colors than usual \citep{2011ApJS..197...19K}. They suggest that low gravity would cause the NIR spectrum to be unusually red. They also eliminate the possibility that the peculiarity arises from binarity, because the estimated companion would have to be a T6.5 dwarf, which would cause a much more peculiar NIR spectrum. Thus, the peculiarity in color is unexplained, and we propose it might be due to cloud-driven variability.


\medskip

\textbf{37) WISE J233527.07+451140.9}

This object was discovered by \citep{2013PASP..125..809T} and visually classified as an L7 dwarf \citep{2016ApJ...833...96L}. \citet{2014ApJ...783..121G} identified this object as an old field candidate with a 97\% probability using BANYAN II, confirmed by BANYAN $\Sigma$ (99.9\% field probability). WISE J2335+4511 is a very red peculiar brown dwarf that does not show signs of weaker K I lines expected in a low gravity object \citep{2013PASP..125..809T}, so it is a part of a class of brown dwarfs that are unusually red objects that are not young.

\subsubsection{Weak Candidates}
\medskip

\textbf{1) 2MASSI J0028394+150141}

2MASSI J0028+1501 is an unusually red mid L dwarf. \citet{2004AJ....127.3553K} suggests that the redness could be due to variability between $5-10\%$. However, no follow-up photometry exists for this object. This source has a 91\% membership probability in Argus according to BANYAN $\Sigma$.


\medskip

\textbf{2) DENIS J025503.3-470049}

DENIS J0255-4700 is an L9 brown dwarf \citep{2014AJ....147...34S}. Variability was not detected in red optical bands $I$ and $R$ in a search by \citet{2005MNRAS.360.1132K}.
\medskip

\textbf{3) 2MASSI J0328426+230205}

2MASSI J0328+2302 is an L9.5 spectral type brown dwarf \citep{2004AJ....127.3553K}. A NIR variability amplitude of  0.43$\pm$0.16\,mag was detected in this object by \citet{2003AJ....126.1006E}, but they were unable to differentiate whether this variability was intrinsic to the brown dwarf or the comparison star. When later observed by \citet{2014ApJ...793...75R}, they found no significant variability in the target over a 3.7\,hr observation. \citet{2010ApJ...710.1142B} noted that this object is overluminous for its spectral type, therefore could be an unresolved binary system.

\medskip

\textbf{4) SDSS J085834.42+325627.7}

This object was first identified by \citet{2006AJ....131.2722C} as a T1 dwarf. Photometric monitoring set variability upper limits of $<0.27\%$ at [3.6] and $<0.64$\% at [4.5] \citep{2015ApJ...799..154M}. SDSS J0858+3256 is unusually red with a large tangential velocity ($V_{tan} = 66\pm 3$\,km s$^{-1}$; \citet{2009AJ....137....1F}) and a 99.9\% probability from BANYAN $\Sigma$ of being a field object. 
\medskip

\textbf{5) SDSSp J132629.82-003831.5}

SDSSp J1326-0038 is an L7 brown dwarf and a candidate member of Argus (40 - 50 Myr) with a $85$\% probability \citep{2015ApJS..219...33G} from BANYAN II. With the updated BANYAN $\Sigma$, however, it has a 100\% probability of being a field object. When observed by \citet{2014ApJ...793...75R}, this object only had a 0.047\% peak-to-peak amplitude in $J$-band.
\medskip

\textbf{6) WISEA J161628.32+062135.1 (PSO J244.1180+06.3598)}

This object was first identified following a crossmatch of the PanSTARRS and WISE surveys as a T0 brown dwarf by \citet{2015ApJ...814..118B}, although it has also been classified as an unusually red L9 brown dwarf. The spectra shows indication of youth because of redder-than-usual colors and triangular $H$-band shape. However, this object has an 99.9\% probability of belonging to the field from BANYAN $\Sigma$.
\medskip

\textbf{7) SDSS J175024.01+422237.8}

SDSS J1750+4222 is a T1 spectral type brown dwarf \citep{2002ApJ...564..466G}. This object has a reported light-curve by \citet{2014ApJ...793...75R}. They found a marginal detection of $1.5\pm 0.3$\% variability amplitude and a $2.7\pm 0.2$\,hr period due to quality concerns with the light
curve.
\medskip

\textbf{8) 2MASS J21522609+0937575}

This object was discovered by \citet{2006AJ....132..891R}. 2MASS J2152+0937 is an equal mass/equal luminosity binary system of L6 spectral type \citep{2008AJ....136.1290R}.

\startlongtable
\begin{longrotatetable}
\begin{deluxetable}{lccccc}
\tablewidth{0pt}
\tablecolumns{6}
\tablenum{4}
\tabletypesize{\scriptsize}
\tablecaption{Variability candidates with spectral types between L7$-$T3\label{tab:candidates}}
\tablehead{
& & & \multicolumn{2}{c}{Variability} & \colhead{Reference}\\
\cline{4-5}
\cline{6-6}
\colhead{Source} &
\colhead{Spectral Type} & 
\colhead{BANYAN $\Sigma$} &
\colhead{Filter} &
\colhead{Amplitude ($\%$ )} &
\colhead{Youth; Variability; Spectral Binary Candidate}
}
\startdata
\cutinhead{\emph{Strong candidates}}
2MASS J00164364+2304262   & T1.0 & 93$\%$ FLD              & \nodata      & \nodata                                          & \nodata ; \nodata ; (2)      \\
2MASS J00310928+5749364   & L9.0 & 96$\%$ CARN             & [3.6]      & 0.35 $\pm$ 0.03                              & (3) ; (3) ; \nodata      \\
2MASS J01020186+0355405   & T0.0 & 98$\%$ FLD              & \nodata      & \nodata                                          & \nodata ; \nodata ; \nodata      \\
2MASS J02062493+2640237   & L8.0 & 67$\%$ FLD              & \nodata      & \nodata                                          & \nodata ; \nodata ; \nodata      \\
WISE J023038.90-022554.0  & L9.0 & 49$\%$ HYA, 34$\%$ CARN*  & \nodata      & \nodata                                          & (1) ; \nodata ; (4)      \\
2MASS J03185403-3421292   & L8.0 & 100$\%$ FLD             & \nodata      & \nodata                                          & \nodata ; \nodata ; \nodata      \\
WISE J032301.86+562558.0  & L7.0 & 100$\%$ FLD             & \nodata      & \nodata                                          & \nodata ; \nodata ; \nodata      \\
$\dagger$ 2MASS J03264225-2102057   & L9.0 & 92$\%$ ABDMG            & [3.6]      & 0.34 $\pm$ 0.04,                             & (3) ; (3) ; \nodata      \\
$\dagger$  2MASS J06420559+4101599   & L8.0 & 86$\%$ ABDMG            & [3.6]      & $2.16 \pm 0.16$                              & (3) ; (3) ; \nodata      \\
2MASS J06522224+4127366   & T0.0 & 100$\%$ FLD*            & \nodata      & \nodata                                          & \nodata ; \nodata ; (5)      \\
$\dagger$  SDSS J075840.33+324723.4  & T2.0 & 94$\%$ CARN*            & J   & 4.8$\pm$0.2                 & (1) ; (6) ; (7)      \\
SDSS J080959.01+443422.2  & L7.0 & 89$\%$ FLD              & [3.6]      & 0.77$\pm$0.06                                & \nodata ; (3) ; \nodata      \\
SDSS J082030.12+103737.0  & T0.0 & 75.5$\%$ FLD*           & \nodata      & \nodata                                          & \nodata ; \nodata ; \nodata      \\
2MASSW J0829570+265510    & L7.0 & 50$\%$ FLD, 50$\%$ ARG*   & \nodata      & \nodata                                          & (1) ; \nodata ; \nodata      \\
SDSSp J083008.12+482847.4 & T0.0 & 100$\%$ FLD             & \nodata      & \nodata                                          & \nodata ; \nodata ; \nodata      \\
SDSSp J083717.22-000018.3 & T1.0 & 100$\%$ FLD             & \nodata      & \nodata                                          & \nodata ; \nodata ; \nodata      \\
2MASSs J0850359+105716    & L7.0 & 51$\%$ FLD, 47$\%$ CARN   & \nodata      & \nodata                                          & (1) ; \nodata ; (8), (9) \\
SDSS J085234.90+472035.0  & T0.0 & 100$\%$ FLD*            & \nodata      & \nodata                                          & \nodata ; \nodata (10)       \\
SDSSp J085758.45+570851.4 & L8.0 & 100$\%$ FLD             & \nodata      & \nodata                                          & \nodata ; \nodata ; \nodata      \\
2MASSI J0859254-194926    & L7.0 & 98$\%$ FLD              & \nodata      & \nodata                                          & \nodata ; \nodata ; \nodata      \\
$\dagger$  2MASSW J0929336+342952    & L8.0 & 69$\%$ ARG*             & \nodata      & \nodata                                          & (1) ; \nodata ; \nodata      \\
$\dagger$  SDSS J093109.56+032732.5  & T0.0 & 90$\%$ CARN*            & J   & 0.041 $\pm$ 0.036                                       & (1) ; (11) ; (12)    \\
2MASS J09553336-0208403   & L8.0 & 56$\%$ FLD, 34$\%$ ABDMG* & \nodata      & \nodata                                          & (1) ; \nodata ; \nodata      \\
2MASS J09560810-1447065   & T0.0 & 100$\%$ FLD*            & \nodata      & \nodata                                          & \nodata ; \nodata ; (2)      \\
SDSS J100711.74+193056.2  & L9.0 & 86$\%$ ARG*             & \nodata      & \nodata                                          & (1) ; \nodata ; \nodata      \\
2MASS J10430758+2225236   & T0.0 & 100$\%$ FLD*            & \nodata      & \nodata                                          & \nodata; \nodata ; (13)      \\
WISE J104915.57-531906.1A & L9.0 & 89$\%$ ARG              & J   & 4.5$\%$                                         & (1) ; (20) ; \nodata                      \\
2MASS J11193254-1137466   & L7.0 & 63.2$\%$ FLD            & [3.6], [4.5] & $0.230^{+0.036}_{-0.0035}$,  $0.453\pm0.037$ & \nodata; (14); \nodata       \\
$\dagger$  2MASS J12010443+5730040   & T0.0 & 71.1$\%$ UMA*           & \nodata      & \nodata                                          & (1) ; \nodata ; \nodata      \\
SDSS J121951.45+312849.4  & T0.0 & 100$\%$ FLD*            & J   & $>2$                              & \nodata ; (15) ; \nodata     \\
SDSSp J125453.90-012247.4 & T2.0 & 100$\%$ FLD             & [3.6], [4.5] & $<0.15$, $<0.36$             & \nodata ; (16) ; \nodata     \\
$\dagger$  2MASSW J1326201-272937    & L7.0 & 88$\%$ CARN             & \nodata      & \nodata                                          & (1) ; \nodata ; \nodata      \\
SDSS J133148.92-011651.4  & T0.0 & 100$\%$ FLD             & \nodata      & \nodata                                          & \nodata ; \nodata ; \nodata      \\
WISE J151314.61+401935.6  & L9.0 & 100$\%$ FLD*            & \nodata      & \nodata                                          & \nodata ; \nodata ; \nodata      \\
SDSS J151506.11+443648.3  & L7.0 & 100$\%$ FLD*            & \nodata      & \nodata                                          & \nodata ; \nodata ; \nodata      \\
SDSS J151643.01+305344.4  & T2.0 & 83$\%$ FLD*             & [3.6], [4.5] & $2.4 \pm 0.2$, $3.1 \pm 1.6$                 & \nodata ; (16) ; \nodata     \\
SDSS J154009.36+374230.3  & L9.0 & 100$\%$ FLD*            & \nodata      & \nodata                                          & \nodata ; \nodata ; \nodata      \\
SDSS J163030.53+434404.0  & L9.0 & 100$\%$ FLD*            & \nodata      & \nodata                                          & \nodata ; \nodata ; \nodata      \\
2MASSW J1632291+190441    & L8.0 & 100$\%$ FLD             & [3.6], [4.5] & $0.42 \pm 0.08$, $0.5 \pm 0.3$               & \nodata ; (16) ; (7)     \\
WISE J164715.57+563208.3  & L8.0 & 66$\%$ FLD              & [3.6]      & 0.47 $\pm$ 0.06                              & (3) ; (3) ; \nodata      \\
WISE J173859.27+614242.1  & L7.0 & 83$\%$ FLD*             & \nodata      & \nodata                                          & \nodata ; \nodata ; \nodata      \\
$\dagger$  WISE J174102.78-464225.5  & L7.0 & 99$\%$ ABDMG            & [3.6]      & $0.35\pm 0.03$                               & (3) ; (3) ; \nodata      \\
WISE J183058.56+454257.4  & L9.0 & 92$\%$ FLD*             & \nodata      & \nodata                                          & \nodata ; \nodata ; (2)      \\
2MASS J18510178+5935040   & L9.0 & 100$\%$ FLD             &  \nodata        &    \nodata                                          &  \nodata ; \nodata ; (4)                    \\
2MASSW J2101154+175658    & L8.0 & 56$\%$ FLD              & \nodata      & \nodata                                          & \nodata ; \nodata ; \nodata      \\
$\dagger$  2MASS J21243864+1849263   & T0.0 & 80$\%$ CARN*            & \nodata      & \nodata                                          & (1) ; \nodata ; \nodata      \\
SDSS J213154.43-011939.3  & L9.0 &             \nodata          & \nodata      & \nodata                                          & \nodata ; \nodata ; \nodata                  \\
2MASS J22153705+2110554   & T0.0 &             \nodata          & J   & 10.7$\pm$0.4                                 & \nodata ; (7) ; (17)     \\
$\dagger$  2MASSW J2244316+204343    & L7.0 & 100$\%$ ABDMG           & J   & 5.5$\pm$0.6                                  & (1) ; (18); \nodata      \\
WISE J232728.74-273056.6  & T0.0 &          \nodata             &  \nodata        &    \nodata                                        & \nodata ; \nodata ; \nodata                      \\
2MASS J23352734+4511442   & L7.0 & 100$\%$ FLD             & \nodata      & \nodata                                          & \nodata ; \nodata ; \nodata     \\
\cutinhead{\emph{Weak candidates}}
$\dagger$   2MASSI J0028394+150141    & L7.0 & 91$\%$ ARG*           & \nodata      & \nodata              & (1) ; \nodata ; \nodata  \\
SDSSp J003259.36+141036.6 & L9.0 & 62$\%$ FLD, 38$\%$ CARN & \nodata      & \nodata              & (1) ; \nodata ; \nodata  \\
WISE J02022929+2305139    & L7.0 & 72$\%$ ARG*           & \nodata      & \nodata              & (1) ; \nodata ; \nodata  \\
DENIS J025503.3-470049    & L9.0 & 100$\%$ FLD           & JHK      & flat             & \nodata ; (19) ; \nodata  \\
2MASSI J0328426+230205    & T0.0 & 100$\%$ FLD           & J   & 0.013            & \nodata ; (11) ; (7) \\
SDSS J073922.26+661503.5  & T2.0 & 100$\%$ FLD*          & \nodata      & \nodata              & \nodata ; \nodata ; \nodata  \\
SDSS J085834.42+325627.7  & T0.0 & 100$\%$ FLD*          & [3.6], [4.5] & $<0.27$, $<0.64$ & \nodata ; (16) ; \nodata \\
SDSSp J132629.82-003831.5 & L7.0 & 100$\%$ FLD           & J   & 0.047            & \nodata ; (11) ; \nodata \\
WISEA J161628.32+062135.1 & L7.0 & 100$\%$ FLD           & \nodata      & \nodata              & \nodata ; \nodata ; \nodata  \\
SDSS J175024.01+422237.8  & T2.0 & 60$\%$ ARG, 40$\%$ FLD* & J   & 2.7$\pm$0.2      & (1) ; (11) ; \nodata \\
2MASS J21522609+0937575   & L7.0 & 100$\%$ FLD*          & \nodata      & \nodata
& \nodata ; \nodata ; \nodata \\
\enddata
\tablerefs{(1) This paper;~(2)~\citet{2015ApJ...814..118B};~(3)~\citet{2022ApJ...924...68V};~(4)~\citet{2013PASP..125..809T};~(5 - \citet{2013ApJ...777...84B};~(6)~\citet{2020AJ....160...38V};~(7)~\citet{2010ApJ...710.1142B};~(8)~\citet{2011AJ....141...70B};~(9)~\citet{2021ApJ...914..124B};~(10)~\citet{2016MNRAS.455.1341M};~(11)~\citet{2014ApJ...793...75R};~(12)~\citet{2015AJ....150..163B};~(13)~\citet{2007AJ....133..439C};~(14)~\citet{2018AJ....155..238S};~(15)~\citet{2014ApJ...782...77B};~(16)~\citet{2015ApJ...799..154M};~(17)~\citet{2019AandA...629A.145E};~(18)~\citet{2019MNRAS.483..480V};~(19)~\citet{2005MNRAS.362..727K};~(20)~\citet{2015ApJ...812..163B}}
\tablecomments{In this table we summarize the literature research on all 62 of our candidates. The spectral type is the spectral type designated by the SpeX spectral classifications. For youth, we provide the BANYAN $\Sigma$ probability of belonging to a young moving group (FLD - field; CARN - Carina Near; ABDMG - AB Doradus; ARG - Argus; UMA - Ursa Major). For objects with previous variability monitoring, we designate the wavelength of the observations and the amplitude of the modulations. Lastly, when available we provide the reference to papers that comment on spectral binary candidacy of any of our objects. Refer to section \ref{sec:bench} for a discussion of notable candidates.}
\tablenotetext{\dagger}{These objects are young based on our classifications which come from both moving group association and spectral indicators of youth in the literature.}
\tablenotetext{*}{These objects do not have measured parallaxes.}
\end{deluxetable}
\end{longrotatetable}

\subsection{Contamination by spectral binaries}

Blended-light spectra can have a binary origin~\citep{2010ApJ...710.1142B,2014ApJ...794..143B} or be the consequence of patchy atmospheric layers at various temperatures, as shown in this study.  In order to estimate the contamination fraction of variability candidates by spectral binaries, we modeled synthetic spectral binary systems and applied our variability detection technique on them. We use our entire spectral library of \nsp~spectra of \sptran~dwarfs 
and applied the~\citet{2010ApJ...710.1142B} technique which is designed to identify T dwarf companions to L7-T3 dwarf primaries. Following this analysis, we remove 87 spectra from 76 sources from our library, which we identify as spectral binary candidates from the~\citet{2010ApJ...710.1142B} technique, many of which have been already confirmed with high resolution follow-up observations. The confirmed and candidate spectral binaries are listed in Table~\ref{tab:SB} for further reference. These spectra, as well as spectra from confirmed variables and our variability candidates, were removed from our sample leaving 147 presumably single, non-variable spectra.

We generated spectral binary templates by interpolating all remaining spectra onto a common wavelength range and then adding every two spectra together. In total, we generated \nsynthbin~spectral binary templates, which were all classified according to the~\citet{2010ApJS..190..100K} prescription. Finally, we apply our variability detection technique on this sample of synthetic spectral binaries. We found that only \sbcontamweak\% of synthetic spectral binaries were selected as weak variability candidates, and an additional \sbcontamstr\% as strong variability candidates, amounting to a total 4.6\% contamination rate of our variability technique by true spectral binaries with similar spectral features as those flagged by our technique. These features could be dependent on the effective temperature of the object, cloud composition, cloud coverage fraction, strength of vertical transport to name a few variables. 

As a complement, we can also provide a rough estimate of the number of candidate and confirmed spectral binaries from our list of 62 variability candidates. We find that 2MASS J0850+1057 is a confirmed closely-separated binary system, whose unresolved spectrum is best fit by a binary template of nearly equal flux components, hence not considered a spectral binary. Six other objects are candidate spectral binaries. However, upon closer inspection, we find that 2 of them are not selected by either the \citet{2010ApJ...710.1142B} or the \citet{2014ApJ...794..143B} spectral binary techniques: WISE J0016+2304 was selected only by indices \citep{2015ApJ...814..118B} and WISE J0230$-$0225 identified as a potential binary by visual inspection \citep{2013PASP..125..809T}. Two other sources, SDSS J0758+3247 and SDSS J0931+0327, do not show any companion on high resolution follow-up (\citealt{2010ApJ...710.1142B} and \citealt{2014ApJ...794..143B}, respectively). Finally, 2MASS J1516+3053 and SDSS J1219+3128 are spectral binary candidates, yet both previously tagged to have unusual atmospheric properties (\citealt{2010ApJ...710.1142B} and \citealt{2019AJ....157..101M}, respectively). This spot check on spectral binaries within our variability candidate list further suggests that our index-based technique will have little contamination from spectral binary systems.

\section{Discussion}\label{sec:discussion}

\subsection{Variability Fractions}

\startlongtable
\begin{deluxetable}{lccc}
\tablewidth{0pt}
\tablecolumns{4}
\tablenum{5}
\tabletypesize{\scriptsize}
\tablecaption{Comparison of variability fractions with the literature\label{tab:varfrac}}
\tablehead{
\colhead{Sample} &
\colhead{This Paper} & 
\colhead{\citet{2014ApJ...793...75R}} &
\colhead{\citet{2015ApJ...799..154M}}
}
\startdata
\cutinhead{L7$-$T3 spectral type range}
Total sources &               270 &                  25 &                  16 \\
Confirmed variables &                15 &                   5 &                   7 \\
Candidates &                62 &                 \nodata &                 \nodata \\
Variability Fraction &   $5_{-1}^{+2}$\% &   $19_{-9}^{+11}$\% & $43_{-17}^{+26}$\% \\
Variability Fraction (inc. cand.\tablenotemark{a}) &  $21\pm3$\% &                 \nodata &                 \nodata \\
\cutinhead{L9$-$T3 spectral type range}
Total sources &               140 &                  15 &                  12 \\
Confirmed variables &                 9 &                   5 &                   4 \\
Candidates &                37 &                 \nodata &                 \nodata \\
Variability Fraction &  $6\pm2$\% &  $33_{-16}^{+21}$\% &  $33_{-18}^{+25}$\% \\
Variability Fraction (inc. cand.\tablenotemark{a}) &  $24^{+5}_{-4}$\% &                 \nodata &                 \nodata \\
\enddata
\tablecomments{For the variability fraction including candidates, we consider both strong and weak candidates.}
\tablenotetext{a}{This fraction assumes that the total number of variables is the number of benchmarks plus 67\% of the variability candidates. These fractions increase to $28_{-3}^{+4}$\% and $33_{-5}^{+6}$\% for L7$-$T3 and L9$-$T3 spectral type ranges, respectively, if 100\% of candidates were to show significant variability amplitudes.}
\end{deluxetable}


In order to assess the success of our technique in predicting variability, we compare variability fractions including our candidates with results from variability surveys from the literature. However, as discussed in Section \ref{sec:bench}, it is important to note that surveys carried out at different wavelengths and with different instruments achieve a wide range of precision and variability amplitudes which differ as a function of wavelength. The comparisons presented here should be interpreted as a rough check of the proportion of  candidate variables obtained using our variability indices.

Within our sample of \nsamp~\sptran~dwarfs (most of which lack photometric monitoring observations), we used \nbench~sources as benchmark variables, leading to a raw variability fraction of $5_{-1}^{+2}\%$ with asymmetrical uncertainties representing the central 68\% of a Poisson distribution (see Table~\ref{tab:varfrac}).~\citet{2014ApJ...793...75R} obtained ground-based light curves in $J$-band for 62 L4–T9 brown dwarfs. From their study, we calculate a raw variability fraction of $19_{-9}^{+11}$\% (5/25) in the \sptran~range that we focus on for this work. Our raw variability fraction is at least $1\sigma$ lower than the one from~\citet{2014ApJ...793...75R} because only a small part of our sample has been followed up with photometric monitoring. 
Similarly, \citet{2015ApJ...799..154M} obtained \emph{Spitzer} light curves in [3.6] and [4.5] bands for 44 L3–T8 dwarfs. Within the \sptran~range, they find 7 variable sources out of 16, leading to a raw fraction of $43_{-18}^{+24}$\%. This fraction is significantly higher than the ground-based fractions from~\citet{2014ApJ...793...75R} and our study because \emph{Spitzer} is more sensitive to low-level variability amplitudes.

A comprehensive photometric monitoring follow-up survey of our candidates is needed to determine the false positive rate of our technique. Among the strong candidates, 14 have previous light curves (including 4 from the new \emph{Spitzer} survey of ~\citet{2022ApJ...924...68V}). Of these, 12 objects have a significant amplitude, whereas 2 do not show significant variability in this wavelength range. From the weak candidates, 4 have previous light curves, none of which exhibit significant variability. Therefore, from this preliminary assessment, it is likely that $67\%$ (12/18) of our candidates will show significant light curve amplitudes after photometric monitoring. Our technique would return almost three times as many significantly variable light curves for L/T transition sources as a randomly-selected sample, using the independent analysis of the near-infrared Brown dwarf Atmosphere Monitoring survey by~\citet{2014ApJ...797..120R} as a comparison ($24^{+11}_{-9}\%$ for L7-T3.5 dwarfs). If this trend continues, 67\% of our \nvarcand~of our variability candidates should show significant variability, and our variability fraction would increase to $21\pm3\%$, which is comparable to the upper limit of~\citet{2014ApJ...793...75R}. However, in order to fully describe our confusion matrix, we would need to homogeneously obtain NIR light curves for every object in our sample, in order to evaluate the effectiveness of our selection technique at identifying those light curves showing significant variability.

~\citet{2014ApJ...793...75R} also carried out a focused analysis on 16 L9–T3.5 dwarfs, out of which 4 had variable light curves, arriving at a fraction of $39^{+16}_{-14}$\% after correcting for their sensitivity to astrophysical signals. Since our spectral type range technically only goes to T3, we recalculated the~\citet{2014ApJ...793...75R} variability fraction for L9-T3 dwarfs (excluding one T3.5 from their sample) and obtain $33_{-16}^{+21}$\%. We find the same fraction for the \citet{2015ApJ...799..154M} sample reduced accordingly to L9-T3, $33_{-18}^{+25}$\%. Within this spectral type range, our raw variability fraction is $6\pm2\%$ for confirmed variables.

\subsection{Atmospheric Structure: Bands, Spots, Inclination Effects}\label{sec:bands}

Variability in a brown dwarf light curve is caused by a heterogeneous cloud coverage of its atmosphere, implying that each longitudinal slice has a different brightness. This difference in brightness in L/T transition objects is probably due to patchy clouds in the atmosphere \citep{2001ApJ...556..872A}. However, not all cloud structures result in variability. If a brown dwarf has zonally or longitudinally symmetric atmospheric cloud features, it will have a flat light curve as the brightness remains the same during the rotation period. Within the solar system, cloud-banding can be seen in Jupiter. Jupiter's banded structure is caused by belts, dark colored bands, and zones, light colored bands. The bands of Jupiter display some variability, but the bulk features remain relatively constant \citep{2019AJ....157...89G}. Atmospheric dynamical simulations show that zonal banding likely occurs for brown dwarfs and giant exoplanets \citep{2021MNRAS.502..678T}. Our technique, which probes the spectra, may be sensitive to blended-light signatures that correspond to cloud patchiness, even if the light curve shows no variability. 

In addition to photometric and spectroscopic studies, brown dwarfs in our candidate sample can also be studied through polarimetry to detect clouds or hazes in their atmosphere if they show flat light-curves. 
Luhman 16 is one of the few accessible targets with polarimetry due to its brightness and distance to Earth. \citet{2020ApJ...894...42M} took \textit{H}-band Very Large Telescope/NaCo linear polarization measurements of the nearby Luhman 16 binary system, and they found that Luhman 16A shows longitudinal cloud bands.  Luhman 16A has shown only mild signs of variability in $iz$ bands through photometric monitoring~\citep{2013ApJ...778L..10B} and in \emph{HST}/WFC3 G102, which covers the peak in the $Y$ band\citep{2015ApJ...812..163B}, although no variability has been detected in longer wavelengths. However, polarimetric measurements from \citet{2020ApJ...894...42M} suggests cloud morphology of banded structures in the NIR. This suggests that the symmetric, banded structure and spin axis inclination of Luhman 16A compound to produce variability amplitudes close to zero~\citep{2015ApJ...798..127B} in the NIR, but we can recover the signature of a blended-light atmosphere (e.g., from different temperature bands) with our empirical technique.

\citet{2017ApJ...842...78V} studied a sample of known \textit{Spitzer Space Telescope} [3.6] and \textit{J}-band variable brown dwarfs, computing the inclination angles of 19 variable brown dwarfs. All objects in the sample with mid-IR variability have an inclination angle greater than 20$^{\circ}$, and objects with \textit{J}-band variability have an inclination angle greater than 35$^{\circ}$. \citet{2017ApJ...842...78V} determined that \textit{J}-band observations could be more affected by inclination, as they probe deeper into the atmosphere. Their study shows that there is a correlation between inclination and variability detection, and even if a brown dwarf does not show variability, this may be due to inclination effects~\citep{2017ApJ...842...78V}: if observed pole-on ($i = 0^{\circ}$), there will be less rotational variations than if observed equator-on ($i = 90^{\circ}$). However, since pole-on orientations are significantly less probable than equator-on~\citep{2010MNRAS.402.1380J}, it is unlikely that most of our variability candidates with future flat light curves would be oriented pole-on. Understanding whether the variability indices technique has a dependence on inclination will be an important step in validating and assessing the technique in future work.

\section{Conclusions}\label{sec:conclusions}

In this paper, we have designed a technique to predict cloud-driven variability in L/T transition brown dwarf atmospheres by studying archival, single-epoch, low-resolution, near-infrared SpeX spectra. We applied the spectral binary indices from \citet{2014ApJ...794..143B} to a sample of \nsamp~\sptran~brown dwarfs with SpeX prism spectra available in the SPL, originally used to highlight differences between unresolved binary systems and single source spectra, in order to identify whether certain indices pinpointed peculiarities in the spectrum caused by variability. We identified regions of the index-index plots where benchmark variables were highly concentrated while minimizing the number of non-variable contaminants. We selected the sources that were repeatedly found in multiple regions of variability and chose them as our candidates for variability study. The candidates found in 10 or 11 of the 11 regions are considered variable candidates, and we recommend them for photometric monitoring follow-up. Our conclusions are as follows:

\begin{enumerate}
    \item We propose a technique to pre-select photometrically variable L7-T3 dwarfs for future monitoring campaigns. Currently there is no method to pre-select which brown dwarfs have higher chances of being variable for monitoring studies. Using this technique can significantly save future resources in next generation telescopes when detecting variability in brown dwarfs.
    \item We have identified \nvarcand~variability candidates from our sample of \nsp~\sptran~integrated light, NIR SpeX spectra following this technique, out of which 50 are entirely new discoveries.
    \item Most of our variability candidates have had peculiar spectra or red colors reported in the literature since their discovery and we suggest that their peculiarities are due to variability. Out of our \nvarcand~candidates, 14 are confirmed young or show kinematic signatures suggestive of youth. Since variability is more prominent among young sources \citep{2017ApJ...842...78V}, our technique might be biased to identify young variable sources. 
    \item We are conducting a ground-based NIR survey to further verify our method to find brown dwarf variables. In a future paper we will demonstrate the efficiency of our novel method.
    \item Flat light curves of our variability candidates might indicate symmetric surface features such as zonal banding or pole-on inclinations.
    \item Based on preliminary photometric monitoring follow up with \emph{Spitzer}, we expect that $\sim67\%$ of our candidates will show a significant variability amplitude.
\end{enumerate}

We have conducted a spectroscopic analysis of photometric variability, finding \nvarcand~unique \sptran~dwarf candidates for variability from their peculiar spectra. If we can indeed identify variability from single-epoch spectra, we can further investigate spectral characteristics of variable brown dwarfs. This will provide further insight into the chemistry of brown dwarf atmospheres and cloud-formation mechanisms.

\begin{acknowledgments}

D.B.G and A.A. acknowledge funding support from the Astrophysics Data Analysis Program of the National Aeronautics and Space Administration under Grant No. 80NSSC19K0532. This research has made heavy use of the VizieR catalogue access tool and SIMBAD database, operated at CDS, Strasbourg, France. The original description of the VizieR service was published in~\citet{2000AandAS..143...23O}, and the SIMBAD astronomical database was published in~\citet{2000AandAS..143....9W}. This publication makes use of data from the SpeX Prism Spectral Libraries, maintained by Adam Burgasser at \url{http://www.browndwarfs.org/spexprism}. We would also like to thank the anonymous reviewer for their careful and thorough comments on our manuscript and their insightful suggestions, which significantly improved our draft. The authors wish to recognize and acknowledge the very significant cultural role and reverence that the summit of Mauna Kea has always had within the indigenous Hawaiian community. We are most fortunate to have the opportunity to conduct observations from this mountain.

\end{acknowledgments}

\vspace{5mm}
\facilities{IRTF(SpeX)}
\software{SPLAT~\citep{2017ASInC..14....7B}}

\appendix

\section{Summary of selection regions for the full sample}
\startlongtable
\begin{longrotatetable}
\begin{deluxetable}{lcccccccccccccc}
\tablewidth{0pt}
\tablecolumns{15}
\tablenum{A.1}
\tabletypesize{\scriptsize}
\tablecaption{Region limits for each index$$-$$index correlation plot.\label{tab:allindices}}
\tablehead{\colhead{Name} & \colhead{SpeX SpT} &  \colhead{Youth} &
\colhead{(1)} &  \colhead{(2)} &  \colhead{(3)} &  \colhead{(4)} &  \colhead{(5)} &  \colhead{(6)} &  \colhead{(7)} &  \colhead{(8)} &  \colhead{(9)} &  \colhead{(10)} &  \colhead{(11)} &  \colhead{Total}}
\startdata
     WISE J000622.67$-$131955.2 &      L8 &      0 &            0 &             0 &            1 &             1 &            0 &            0 &            0 &              0 &             0 &              0 &             0 &      2 \\
       WISE J00062785+1857288 &      L7 &      0 &            0 &             0 &            0 &             0 &            0 &            0 &            1 &              0 &             0 &              1 &             1 &      3 \\
      2MASS J00132229$-$1143006 &      T3 &      0 &            0 &             0 &            0 &             0 &            0 &            0 &            0 &              0 &             0 &              0 &             0 &      0 \\
      2MASS J00133470+1109403 &      L7 &      0 &            0 &             0 &            1 &             1 &            0 &            0 &            1 &              0 &             0 &              1 &             1 &      5 \\
      2MASS J00150206+2959323 &      L8 &      0 &            0 &             0 &            0 &             0 &            0 &            0 &            0 &              0 &             0 &              0 &             0 &      0 \\
        PSO J004.1834+23.0741 &      T0 &      0 &            1 &             1 &            1 &             1 &            1 &            1 &            1 &              1 &             1 &              1 &             1 &     11 \\
       WISE J00164397+2304265 &      T1 &      0 &            1 &             1 &            1 &             1 &            1 &            1 &            1 &              1 &             1 &              1 &             1 &     11 \\
        PSO J004.7148+51.8918 &      L7 &      0 &            1 &             1 &            1 &             1 &            0 &            1 &            1 &              1 &             0 &              1 &             1 &      9 \\
\enddata
\end{deluxetable}
\tablecomments{Order of columns (1) H$_2$O$-$J vs. CH$_4-$H, (2) H$_2$O$-$J vs. H$-$bump, (3) CH$_4-$J vs. CH$_4-$H, (4) CH$_4-$J vs. H$-$bump, (5) CH$_4-$H vs. H$_2$O$-$K, (6) CH$_4-$H vs. CH$_4-$K, (7) CH$_4-$H vs. H$-$dip, (8) CH$_4-$H vs. J$-$curve, (9) H$_2$O$-$K vs. H$-$bump, (10) H$-$dip vs. K$-$slope, (11) H$-$dip vs. H$-$bump}
\end{longrotatetable}

\section{Candidate and confirmed spectral binaries in the sample}
\startlongtable
\begin{deluxetable}{lcccccc}
\tablewidth{0pt}
\tablecolumns{7}
\tablenum{A.2}
\tabletypesize{\scriptsize}
\tablecaption{L7$-$T3 dwarf spectral binary candidates from our index-selection technique.\label{tab:SB}}
\tablehead{
\colhead{Source} &
\colhead{SpeX SpT} & 
\colhead{$J$ mag} & 
\colhead{$K$ mag} & 
\colhead{Total} & 
\colhead{SNR} & 
\colhead{Reference\tablenotemark{a}}}
\startdata
\cutinhead{Confirmed Spectral Binaries}
    2MASS J04473039-1216155  &    T1.0 &   16.48$\pm$0.11 &   15.55$\pm$0.22 &      5 &   10.10  &  3;1;3 \\ 
    2MASS J05185995$-$2828372 &    T2.0 &   15.978$\pm$0.1 &  14.162$\pm$0.07 &      6 &   56.00 &  13;13;5\\
    WISE J061135.13$-$041024.0 &    T0.0 &  15.489$\pm$0.06 &  14.221$\pm$0.07 &      3 &   23.00 &  20;18;18 \\
    SDSS J080531.84+481233.0 &    L7.0 &  14.734$\pm$0.04 &  13.444$\pm$0.04 &      3 &   23.00 &  12;7;15\\
    Gl 337CD &    L9.0 &  15.512$\pm$0.08 &  14.043$\pm$0.06 &      2 &   84.00 &  9;1;15 \\ 
    SDSS J102109.69-030420.1 &    T3.0 &  16.253$\pm$0.09 &  15.126$\pm$0.17 &      6 &   31.00 &  5;18;15\\
    WISE J104915.57$-$531906.1A &    L9.0 &  11.511$\pm$0.03 &   9.559$\pm$0.03 &      2 &  133.00 &  10;10;23\\
    2MASS J11061191+2754215 &    T2.0 &  14.879$\pm$0.03 &  13.788$\pm$0.04 &      4 &   78.00 &  24;9;30\\
\enddata
\tablerefs{(1) This paper;~(2)~\citet{2013ApJ...777...84B};~(3)~\citet{2015ApJ...814..118B};~(4)~\citet{2004AJ....127.2856B};~(5)~\citet{2006ApJ...637.1067B};~(6)~\citet{2006ApJS..166..585B};~(7)~\citet{2007AJ....134.1330B};~(8)~\citet{2008ApJ...681..579B};~(9)~\citet{2010ApJ...710.1142B};~(10)~\citet{2013ApJ...772..129B};~(11)~\citet{2015AJ....150..163B};~(12)~\citet{2006AJ....131.2722C};~(13)~\citet{2004ApJ...604L..61C};~(14)~\citet{2011AJ....142...77D};~(15)~\citet{2012ApJS..201...19D};~(16)~\citet{2012ApJ...757..100D};~(17)~\citet{2013MNRAS.430.1171D};~(18)~\citet{2014AJ....148....6G};~(19)~\citet{2011ApJ...732...56G};~(20)~\citet{2011ApJS..197...19K};~(21)~\citet{2015AJ....150..182K};(22)~\citet{2017AJ....154..112K}~(23)~\citet{2013ApJ...767L...1L};~(24)~\citet{2007AJ....134.1162L};~(25)~\citet{2010ApJ...722..311L};~(26)~\citet{2007ApJ...654..570L};~(27)~\citet{2014ApJ...787..126L};~(28)~\citet{2006ApJ...647.1393L};~(29)~\citet{2013ApJS..205....6M};~(30)~\citet{2013AandA...560A..52M};~(31)~\citet{2008ApJ...676.1281M};~(32)~\citet{2016ApJ...830..144R};~(33)~\citet{2006ApJ...639.1114R};~(34)~\citet{2009AJ....137..304S};~(35)~\citet{2017AJ....153..196S};~(36)~\citet{2016ApJ...817..112S};~(37)~\citet{2013PASP..125..809T};~(38)~Burgasser et al. inprep.}
\tablenotetext{a}{Spectrum; Spectral Binary Candidate; Confirmed Binarity/Variability}
\end{deluxetable}

\bibliographystyle{aastex}{}
\bibliography{main.bib}

\end{document}